\documentclass[superscriptaddress,aps,preprintnumbers,amsmath,amssymb,prd,nofootinbib,preprint,longbibliography]{revtex4-1}
\pdfoutput=1

\usepackage{graphicx,array}
\usepackage{color}
\usepackage{cancel}
\usepackage{latexsym}
\usepackage{amsthm}
\usepackage{amsmath}
\usepackage{amssymb}
\usepackage{hyperref} 
\usepackage{bbold}
\usepackage{mathtools}
\usepackage{enumitem}

\numberwithin{equation}{section}

\makeatletter
\renewcommand{\p@subsection}{}
\renewcommand{\p@subsubsection}{}
\makeatother

\def\simgt{\mathrel{\lower2.5pt\vbox{\lineskip=0pt\baselineskip=0pt
           \hbox{$>$}\hbox{$\sim$}}}}
\def\simlt{\mathrel{\lower2.5pt\vbox{\lineskip=0pt\baselineskip=0pt
           \hbox{$<$}\hbox{$\sim$}}}}

\newcommand{\be}{\begin{equation}}
\newcommand{\ee}{\end{equation}}
\newcommand{\bea}{\begin{eqnarray}}
\newcommand{\eea}{\end{eqnarray}}

\newcommand{\GeV}{\textrm{ GeV}}
\newcommand{\TeV}{\textrm{ TeV}}
\newcommand{\gsim}{\lower.7ex\hbox{$\;\stackrel{\textstyle>}{\sim}\;$}}
\newcommand{\lsim}{\lower.7ex\hbox{$\;\stackrel{\textstyle<}{\sim}\;$}}

\newcommand{\MPl}{M_{\rm Pl}}

\begin{document}

\title{Cosmic Perturbations from a Rotating Field}

\author{Raymond T. Co}
\affiliation{William I. Fine Theoretical Physics Institute, School of Physics and Astronomy, University of Minnesota, Minneapolis, MN 55455, USA}
\author{Keisuke Harigaya}
\affiliation{Theoretical Physics Department, CERN, Geneva, Switzerland}
\author{Aaron Pierce}
\affiliation{Leinweber Center for Theoretical Physics, Department of Physics, University of Michigan, Ann Arbor, MI 48109, USA}

\date{\today}

\begin{abstract}
Complex scalar fields charged under approximate $U(1)$ symmetries appear in well-motivated extensions of the Standard Model. One example is the field that contains the QCD axion field associated with the Peccei-Quinn symmetry; others include flat directions in supersymmetric theories with baryon, lepton, or flavor charges.  These fields may take on large values and rotate in field space in the early universe. The relevant approximate $U(1)$ symmetry ensures that the angular direction of the complex field is light during inflation and that the rotation is thermodynamically stable and is long-lived. These properties allow rotating complex scalar fields to naturally serve as curvatons and explain the observed perturbations of the universe. The scenario imprints non-Gaussianity in the curvature perturbations, likely at a level detectable in future  large scale structure observations. The rotation can also explain the baryon asymmetry of the universe without producing excessive isocurvature perturbations. 
\end{abstract}

\preprint{UMN-TH-4113/22, FTPI-MINN-22-04, CERN-TH-2022-007, LCTP-22-02}

\maketitle

\tableofcontents

\newpage

%%%%%%%%%%%%%%%%%%%%%%%%%%%%
\section{Introduction}
\label{sec:intro}
%%%%%%%%%%%%%%%%%%%%%%%%%%%%

In the simplest inflationary cosmology, an inflaton field is not only responsible for solving the horizon and flatness problems, but also for the adiabatic perturbations that seed the structure of the universe.  However, these perturbations could arise from fluctuations imprinted in another field that is light during the inflationary epoch,  the so-called curvaton~\cite{Enqvist:2001zp,Lyth:2001nq,Moroi:2001ct}. The curvaton must be also long-lived so that it may come to dominate the energy density of the universe; this ensures the non-Gaussianity of cosmic perturbations is not too large~\cite{Lyth:2002my}.  Also, for a curvaton scenario to be successful, the observed baryon asymmetry and dark matter must originate from the curvaton itself or the entropy created from it; otherwise too-large  matter isocurvature perturbations are produced~\cite{Lyth:2002my}.

But what is the identity of the curvaton? This question is pressing given the non-minimality of the curvaton scenario, coupled with the required special properties of the curvaton, namely its lightness and longevity. Here, we show that a rotating complex scalar field associated with an approximate $U(1)$ symmetry can naturally play the role of a curvaton, because of the lightness of the angular direction and the longevity of the rotation. Complex scalar fields associated with approximate $U(1)$ symmetries are ubiquitous in theories beyond the Standard Model (SM). In the axion solution to the strong CP problem~\cite{Peccei:1977hh,Peccei:1977ur,Weinberg:1977ma,Wilczek:1977pj}, a complex scalar field associated with the Peccei-Quinn (PQ) symmetry is introduced, and the PQ symmetry is explicitly broken by the QCD anomaly, so the symmetry is only approximate. In supersymmetric theories, scalar superpartners of SM particles are introduced, and combinations of them are charged under approximate symmetries such as baryon number plus lepton number $(B+L)$ or flavor symmetries. 

Because these symmetries are only approximate, it is plausible that they may also be explicitly broken by a higher-dimensional potential for our complex field $P$. Then rotations of $P$, which correspond to a non-zero $U(1)$ charge, may be induced by a mechanism employed in  Affleck-Dine baryogenesis~\cite{Affleck:1984fy,Dine:1995uk,Dine:1995kz}; namely, a ``kick" may be given to the angular direction by the higher-dimensional explicit breaking. A large early-universe field value for $P$ may result from a negative Hubble-induced mass term; such a large field value enhances the importance of the $U(1)$-breaking term. The angular direction of $P$ remains light during inflation because of the approximate $U(1)$ symmetry and thus obtains quantum fluctuations~\cite{Mukhanov:1981xt,Hawking:1982cz,Starobinsky:1982ee,Guth:1982ec,Bardeen:1983qw}. These fluctuations lead to fluctuations in the $U(1)$ charge induced by the kick. 

The rotation initially contains both angular and radial motion, but once $P$ is thermalized via interactions with the thermal bath, the radial motion dissipates, and the rotation becomes circular~\cite{Co:2019wyp}. Upon thermalization, a small fraction of the charge is transferred into a particle-antiparticle asymmetry in the thermal bath, but it is free-energetically favored to keep almost all of the charge in the form of rotation~\cite{Laine:1998rg,Co:2019wyp}. This thermodynamic stability ensures the longevity of the rotation.

We assume that $P$ has a nearly quadratic potential. This is indeed the case for the flat directions in the Minimal Supersymmetric Standard Model (MSSM) (see~\cite{Gherghetta:1995dv} for a survey of MSSM flat directions) and for a class of PQ symmetry breaking fields. With this assumption, the energy density of the rotation decreases due to cosmic expansion as matter, so this energy density can eventually exceed that of radiation. The present-day entropy of the Universe can then be created by the (partial) washout of the $U(1)$ charge, and the fluctuations of the $U(1)$ charge create curvature perturbations; the rotation functions as a curvaton.

Similar to the conventional curvaton that arises from an oscillating field \cite{Lyth:2002my}, this scenario predicts an observable amount of local non-Gaussianity in the curvature perturbations. In most of the parameter space, the predicted local non-Gaussianity is allowed by the current CMB limit~\cite{Planck:2019kim} but can be detected by future observations of small scale structure~\cite{Alvarez:2014vva,Dore:2014cca}.

Our scenario can somewhat mitigate the $\eta-$problem; generically, there is no reason that the $\eta$ parameter of the inflaton is small~\cite{Ovrut:1983my,Holman:1984yj,Goncharov:1983mw,Coughlan:1984yk}. In a curvaton scenario, a sufficiently small $\eta$ associated with the inflaton potential is still required so that slow-roll inflation can occur. But in a standard inflationary scenario, even smaller values of $\eta$ must hold to reproduce the observed small tilt of the perturbation spectrum. And while anthropic considerations might explain the $\eta$ necessary for inflation, it is hard to understand how they might motivate even smaller values of $\eta$. (See, however,~\cite{Tegmark:2004qd,Masoumi:2016eag,Chiang:2018dju}.) Consistency with the observed spectral tilt will require the $\eta$ parameter of the curvaton itself to be smaller than $\mathcal{O}(0.01)$, but that can be understood by the approximate $U(1)$ symmetry in our scenario.

The washout required in our scenario can occur from explicit breaking of the $U(1)$ symmetry.  In the case that the complex scalar is protected by a $B+L$ symmetry, the washout can be due to the weak sphaleron process. If the relevant symmetry is the PQ symmetry, as would be the case for the axion, strong sphalerons could do the job. (As we will discuss, the washout actually involves a combination of processes, so that no remaining linear combination of approximate symmetries  protects the rotation.) The washout rate is suppressed by the square of the ratio between the temperature and the radius of the rotation~\cite{Co:2019wyp}, so it is suppressed at early times but can become effective as the radius shrinks by redshifting.

In this paper, we discuss two well-motivated examples mentioned above, flat directions in the MSSM and a PQ symmetry breaking field. For the former, washout easily occurs since the radius of the rotation can become arbitrarily small, and the washout rate becomes larger than the Hubble expansion rate.  For the latter, whether or not the washout is effective is non-trivial since the radius of rotation cannot be smaller than the axion decay constant. Also, in the time slice where the PQ charge is uniform, the axion field itself is not uniform.  A consequence is that domain walls without boundaries may be produced from the fluctuation of the axion field. Such domain walls could lead to an unacceptable cosmology. The fluctuations can also produce matter isocurvature perturbations through misalignment \cite{Preskill:1982cy, Dine:1982ah,Abbott:1982af} axion dark matter production.  We argue how these problems can be avoided by the resultant symmetry restoration as arising from either thermal trapping after the washout or an era of parametric resonance in the early stage of the rotation. 

The rotation has implications for the baryon asymmetry and dark matter. The $U(1)$ charge of the MSSM flat directions can be converted into a baryon asymmetry through $B-L$ violating interactions~\cite{Chiba:2003vp,Takahashi:2003db,Domcke:2020kcp,Co:2020jtv,Domcke:2020quw}. The PQ charge in the axion rotation can be converted into a baryon asymmetry (axiogenesis)~\cite{Co:2019wyp,Co:2020xlh,Co:2020jtv,Harigaya:2021txz,Chakraborty:2021fkp,Kawamura:2021xpu,Co:2021qgl}. The kinetic energy of the axion rotation may also be the source of the axion dark matter density (kinetic misalignment)~\cite{Co:2019jts,Co:2020dya}.  Since the rotation dominates the energy density of the universe, and the entropy of the universe almost entirely comes from it, baryon and dark matter may be produced from the $U(1)$ charge without producing matter isocurvature perturbations. In this scenario, three essential ingredients of the universe---the dark matter, the baryon asymmetry, and the density fluctuations---can all originate from the rotation of a complex field. 

Our mechanism requires a nearly quadratic potential for the radial direction of the complex field so that the rotation can dominate the universe. This is naturally realized in supersymmetric theories. Motivations of supersymmetry cannot be stressed too much; it is the unique extension of the space-time symmetry, achieves precise gauge coupling unification, relaxes the hierarchy problem, and provides the lightest supersymmetric particle as a dark matter candidate. We therefore perform our analysis in the context of supersymmetric theories, but the key idea can be applied to non-supersymmetric theories if the potential for the radial direction is nearly quadratic. 

Curvaton models that share some similarities with the present scenario have been investigated in the literature. For example, axion-like fields were proposed as curvaton candidates in order to understand the lightness of the curvaton~\cite{Dimopoulos:2003az,Kawasaki:2011pd,Kawasaki:2012gg,Kobayashi:2020xhm}.  However, in these works, the motion of the axion fields was not rotation but rather oscillation (as driven by the vacuum potential).  For the axion-like curvaton field to dominate and decay, the required mass is much larger than that of the QCD axion, so the most well-motivated axion cannot do the job in that case. MSSM flat directions acting as curvatons are discussed in Refs.~\cite{Enqvist:2002rf,Enqvist:2003mr,Kasuya:2003va,Hamaguchi:2003dc}. The possibility of producing cosmic perturbations from the fluctuation of the angular direction of a rotating complex field  is discussed in Refs.~\cite{McDonald:2003jk,Riotto:2008gs}. They, however, do not take into account the thermodynamical stability of the rotation, and in these scenarios the energy density of the angular mode is at the most comparable to that of the radial mode. This not only leads to non-Gaussianity that is too large unless the energy densities are comparable, but it also makes baryogenesis from the charge in the rotation impossible, as correlated baryon isocurvature perturbations become too large. Ref.~\cite{Harigaya:2019uhf} considers the generation of curvature perturbations from the decay of Q-balls created from the rotation. In this case, the energy density dominantly comes from the angular mode, and the non-Gaussianity is not too large. However, because of fluctuations in the Q-ball lifetimes, if dark matter or baryon are produced from Q-balls, correlated matter isocurvature perturbations are generically produced. We will discuss the conditions for avoiding Q-ball formation in our scenario.

This paper is organized as follows. The dynamics of rotating fields are reviewed in Sec.~\ref{sec:rotating}. We compute the spectrum of the curvature perturbations in Sec.~\ref{sec:perturbation}. The case of the PQ symmetry breaking field is discussed in Sec.~\ref{sec:axion}. Discussion including comments on baryogenesis from the rotation is given in Sec.~\ref{sec:Disc}. In the Appendices, we discuss the thermalization of rotating MSSM flat directions, provide details on the numerical computation of the charge density of the rotating field, and discuss cosmic strings that rotate in field space. 

%%%%%%%%%%%%%%%%%%%%%%%%%%%%
\section{Rotating field}
\label{sec:rotating}
%%%%%%%%%%%%%%%%%%%%%%%%%%%%
In this section, we discuss the dynamics of a rotating complex scalar field $P$, which has a radial direction $S$ and an angular direction $\theta$,
\begin{align}
    P = \frac{1}{\sqrt{2}}S e^{i \theta}.
\end{align}
We assume that $P$ has a (nearly) quadratic $U(1)$ symmetric potential. As we will see, this allows the rotation to dominate the energy density of the universe and thus produce the curvature perturbations of the universe without generating too-large non-Gaussianity.  Since the assumption is naturally realized in supersymmetric theories, we discuss the dynamics of rotations in the context of supersymmetric theories.

\subsection{Initiation of the rotation}
The $U(1)$ symmetry may be explicitly broken by a higher-dimensional operator in the superpotential%
\footnote{$U(1)$ symmetry breaking in the K\"ahler potential can also initiate the rotation~\cite{Harigaya:2016hqz}. For this case, however, the field value of $P$ must be around the cutoff scale for the initiation of the rotation to be effective, and theoretical control is lost. In particular, the angular direction can obtain a mass as large as the Hubble scale and may not obtain fluctuations.}
that is negligible at the vacuum,
\begin{align}
\label{eq:PQV}
    W = \frac{P^{n+1}}{M^{n-2}},
\end{align}
where $M$ is a dimensionful parameter. The effect of such explicit $U(1)$ breaking may be enhanced in the early universe, and it may initiate the rotation of $P$ by the Affleck-Dine mechanism~\cite{Affleck:1984fy,Dine:1995uk,Dine:1995kz}, which we review below. 

The scalar potential of $P$ is given by
\begin{align}
\label{eq:potential}
    V = \frac{(n+1)^2}{M^{2n-4}}|P|^{2n} + m_S^2 |P|^2 - c_H H^2 |P|^2 + \left( (n-2) a_{\cancel{U(1)}} m_{S} \frac{P^{n+1}}{M^{n-2}} + {\rm h.c.}\right),
\end{align}
where the first term is from the $F$ term of $P$, the second term is from the supersymmetry breaking at the vacuum, the third term is from the supersymmetry breaking in the early universe,  the so-called Hubble-induced mass, and the last term is from $R$ symmetry breaking at the vacuum and the superpotential in Eq.~(\ref{eq:PQV}), the so-called $A$ term. The constant $a_{\cancel{U(1)}}$ is expected to be ${\mathcal O }(m_{3/2}/m_S)$, with $m_S$  the mass of the radial mode of $P$ and $m_{3/2}$ the gravitino mass. If the soft mass of $S$ is given by gravity mediation, $a_{\cancel{U(1)}} = {\mathcal O}(1)$ is expected.

Assuming a positive $c_H \gsim \mathcal{O}(1)$ during inflation, the radial direction of $P$ is fixed at a large field value where the first and the third terms balance,
\begin{align}
\label{eq:Pinf}
    |P| = \left[ \frac{c_H^{1/2} H_{\rm inf} M^{n-2}}{n^{1/2}(n+1)} \right]^{\frac{1}{n-1}},
\end{align}
where $H_{\rm inf}$ is the Hubble scale during inflation. For $H_{\rm inf} \gg m_S$, the mass of the angular direction given by the last term in Eq.~(\ref{eq:potential}) is much smaller than $H_{\rm inf}$, and the angular direction obtains fluctuations. As we will see in the next section, the fluctuations can produce the observed curvature perturbations of the Universe.

After inflation, $P$ is no longer fixed at the field value in Eq.~\eqref{eq:Pinf} but evolves as the Hubble scale changes. During this period, the inflaton oscillates around the minimum of its potential and eventually decays into radiation. We assume that the thermal potential of $P$ is negligible when its rotation is initiated, which is the case when the field value of $P$ is large and/or the reheat temperature following inflation is small enough. We also assume that after inflation, $c_H$ remains positive, or negative but is smaller than $\mathcal{O}(1)$. Then the radial direction follows an attractor solution as long as $m_S \ll H$~\cite{Dine:1995kz,Harigaya:2015hha},  
\begin{align}
\label{eq:Pafter}
    |P| = \left[\frac{ 3(1+w)H M^{n-2}}{2 n^{1/2} (n+1)} \left(\frac{4 c_H}{9(1+w)^2} + \frac{2}{(n-1)(1+w)} - \frac{n}{(n-1)^2}\right)^{ \scalebox{0.9}{$\frac{1}{2}$} } \right]^\frac{1}{n-1},
\end{align}
where $w = 1/3$ for radiation domination and $0$ for matter domination.

When $3H$ becomes comparable to $m_S$, oscillations in $P$ are induced by the vacuum potential. At the same time, the last term in Eq.~(\ref{eq:potential}) gives a kick to $P$ in the angular direction, and $P$ begins to rotate around the origin. Because of cosmic expansion, the field value of $P$ decreases, and shortly after the beginning of the rotation, the explicit $U(1)$ breaking potential becomes negligible. $P$ then continues to rotate while preserving the angular momentum in field space (the $U(1)$ charge density) up to the dilution by cosmic expansion,
\begin{align}
    n_{\theta} = i (\dot{P}P^*-  \dot{P}^* P) = - \dot{\theta} S^2 \propto R^{-3},
\end{align}
where $R$ is the scale factor of the universe.
The above dynamics can be explicitly seen from the equation for $n_\theta$,
\begin{align}
\label{eq:nthetaeom}
    \dot{n}_\theta + 3H n_\theta = \frac{\partial V}{\partial \theta},
\end{align}
where the right-hand side is responsible for the kick.

It is convenient to normalize the charge density by the entropy density $s$,
\begin{align}
    Y_\theta \equiv \frac{n_\theta}{s},
\end{align}
since $Y_\theta$ remains constant as long as entropy is conserved. We may evaluate $Y_{\theta}$ in the following way. The potential gradient in the angular direction is about $a_{\cancel{U(1)}}$ times that in the radial direction when the rotation is initiated at $3H \simeq m_S$; see Eqs.~(\ref{eq:potential}) and (\ref{eq:Pafter}). Therefore, the kick induces $\dot\theta \simeq a_{\cancel{U(1)}} m_S$. Here we assume that the initial angle is not close to the minimum nor the maximum of the potential generated by the explicit breaking term. Then the $U(1)$ charge density at the time of the initiation of the rotation is about $a_{\cancel{U(1)}} m_S S_i^2$, where $S_i$ is the field value of $S$ when the rotation begins. The resultant $Y_\theta$ is given by
\begin{align}
\label{eq:Ytheta}
    Y_\theta \simeq
    \begin{cases}
    50 a_{\cancel{U(1)}} 
    \left(\frac{100\TeV}{m_S}\right)^{\scalebox{0.9}{$\frac{1}{2}$} } 
    \left(\frac{S_i}{10^{16} \GeV}\right)^2 
    \left(\frac{200}{g_*}\right)^{\scalebox{0.9}{$\frac{1}{4}$} }
    & : T_R \gtrsim 10^{11} \GeV 
    \left(\frac{m_S}{100 \TeV}\right)^{ \scalebox{0.9}{$\frac{1}{2}$} } 
    \left(\frac{200}{g_*}\right)^{\scalebox{0.9}{$\frac{1}{4}$} }\\ 
    4 a_{\cancel{U(1)}} 
    \left(\frac{T_R}{10^{10}\GeV}\right) 
    \left(\frac{100 \TeV}{m_S}\right) 
    \left(\frac{S_i}{10^{16} \GeV}\right)^2 
    & : T_R \lesssim 10^{11} \GeV 
    \left(\frac{m_S}{100 \TeV}\right)^{ \scalebox{0.9}{$\frac{1}{2}$} }
    \left(\frac{200}{g_*}\right)^{\scalebox{0.9}{$\frac{1}{4}$} }
    \end{cases}.
\end{align}
Here $T_R$ is the reheat temperature after inflation, and we assume that no entropy is produced after the completion of the reheating by the inflaton. (The case with entropy production from the radial mode of $P$ is discussed in Appendix~\ref{sec:thermalization}.) The first (second) expression corresponds to the case where the rotation begins  after (before) the completion of the reheating. $Y_\theta$ for the second case can be estimated by first computing the redshift-invariant ratio $n_\theta/\rho_{\rm inf}$ with $\rho_{\rm inf}$ the inflaton energy density.  This ratio can be computed straightforwardly at the onset of the rotation and can then be converted to $n_\theta/s$ at the completion of reheating.

\subsection{Evolution of the rotation}
\label{sec:evolution}

The evolution of the rotation after its  initiation is discussed in Ref.~\cite{Co:2019wyp}. At first, the rotation contains both radial and angular motion. The rotation is thermalized via its interaction with the thermal bath. The radial mode is dissipated, but because of $U(1)$ charge conservation, the angular mode is nearly unaffected; while some $U(1)$ charge is transferred into particle-antiparticle asymmetries in the thermal bath,  it is free-energetically favorable to keep almost all of the charge in the form of rotation as long as the $U(1)$ charge density is larger than $m_S T^2$~\cite{Laine:1998rg,Co:2019wyp}. This can be seen in the following way. The chemical potential of the rotation is $\dot{\theta}$, so in equilibrium, the chemical potential of the $U(1)$ charge asymmetry of the bath is also $\mathcal{O}(\dot{\theta})$.  The result is that the bath asymmetry is $\mathcal{O}(\dot{\theta} T^2)$, which is suppressed relative to the charge in the rotation $n_{\theta} = -\dot{\theta} S^2$---this proves the thermodynamic stability of the rotation. The angular motion is retained; the result is nearly circular motion without ellipticity. Eventually, the charge can be washed out by explicit breaking of the $U(1)$ symmetry, as discussed in Sec.~\ref{sec:washout}. Here we discuss the evolution well before the washout.

When the motion is circular, the equation of motion of $S$ requires that
\begin{align}
\label{eq:dtheta}
    \dot{\theta}^2 = \frac{V'(S)}{S},
\end{align}
which is as large as $m_S^2$. Charge conservation requires $n_\theta \propto R^{-3}$. Thus, for the quadratic potential at hand, the evolution of $\dot{\theta}$, $S$, and the energy density of the rotation $\rho_{\theta}$ is
\begin{align}
    \dot{\theta} \propto R^0,~~
    S^2 \propto R^{-3},~~
    \rho_\theta \propto R^{-3}.
\end{align}
The energy density of the rotation redshifts more slowly than radiation ($R^{-3}$ vs.~$R^{-4}$), and it can come to dominate the energy density of the universe. The temperature at which this transition from radiation to matter domination occurs is
\begin{align}
\label{eq:TRM}
    T_{\rm RM} & = \frac{4}{3}m_S Y_{\theta} 
    \simeq  130 \TeV \left(\frac{m_S}{\rm TeV}\right) \left(\frac{Y_{\theta}}{100}\right).
\end{align}
Dynamics of the rotation is extensively investigated in~\cite{Co:2019wyp,Co:2019jts,Co:2020jtv,Co:2021rhi,Kawamura:2021xpu,Co:2021lkc} for a PQ symmetry breaking field---including the thermalization of the radial mode and possible entropy production from it---and it is found that the circular rotation can indeed dominate the universe~\cite{Co:2021lkc} for a wide range of interesting parameters. In Appendix~\ref{sec:thermalization}, we demonstrate that rotations of MSSM flat directions can similarly come to dominate. There, we also discuss the case where the rotation dominates the universe before the completion of the thermalization.

Note that Eq.~(\ref{eq:dtheta}) relates $\dot{\theta}$ and $S$, so one variable is sufficient to specify the state of the rotation. As we will see, $n_{\theta}$ is a convenient variable to compute the curvature perturbations produced from the rotation.

\subsection{Q-ball formation and parametric resonance}
\label{sec:Qball_PR}

In the above discussion, we assumed that the rotation is coherent. We now discuss two processes that can create inhomogeneities in $P$: Q-ball formation and parametric resonance.  We find that the possibility of Q-ball formation can place non-trivial constraints on the potential of the flat direction.   Parametric resonance, on the other hand, is not problematic but will play an important role in the case where the rotation of the PQ field is the curvaton. 

A purely quadratic potential of $S$ is at the boundary of the condition for Q-ball formation, so small perturbations to the potential could lead to Q-balls. Indeed, the potential of $S$ is not purely quadratic but receives a logarithmic correction $K m_S^2|P|^2 {\rm ln}|P|^2$ because of quantum corrections. If $K<0$, Q-balls may form~\cite{Coleman:1985ki,Kusenko:1997zq,Kusenko:1997si,Kasuya:1999wu,Dine:2003ax}. If Q-balls come to dominate the energy density of the universe, they will induce too-large isocurvature perturbations if the baryon asymmetry is produced by the rotation, as explained in Sec.~\ref{sec:perturbation}.  (See~\cite{Harigaya:2019uhf} for the case of Q-ball formation with naturally compensated isocurvature perturbations.) For the PQ symmetry-breaking field, $K>0$ can be achieved by the Yukawa coupling between $P$ and other fields, such as KSVZ (Kim-Shifman-Vainshtein-Zakharov)~\cite{Kim:1979if,Shifman:1979if} quarks, that have positive scalar soft mass squared. For MSSM flat directions, requiring $K>0$ restricts the possible choice of the flat directions depending on the mass spectrum, as the quantum corrections from the gaugino masses and gauge interactions make negative contributions to $K$. For scalar masses $m_S \sim m_{\lambda}$ with $m_{\lambda}$ the gaugino masses, if the flat direction contains $Q_3$ or $\bar{u}_3$, the corrections from the top Yukawa and scalar masses can dominate $K$, ensuring $K>0$. For large ${\rm tan} \beta$, $\bar{d}_3$, $\bar{e}_3$, and $L_3$ also work. For scalar masses $m_S > 10m_{\lambda}$ (natural because of the chiral symmetry that can suppress gaugino masses) as indeed occurs in the without-singlet scenario~\cite{Giudice:1998xp}, even flat directions that exclusively include the first and the second generations can have $K>0$ since the logarithmic correction to the $S$ potential from the gaugino masses are suppressed by $(m_{\lambda}/m_S)^2$, while the two-loop corrections from the gauge interactions and scalar masses~\cite{Martin:1993zk} can generate positive $K= 10^{-4}$ and $10^{-3}$ for non-colored and colored scalars, respectively. But even in the case where $K>0$, thermal corrections could, in principle, also lead to production of Q-balls. However, as is shown in Appendix~\ref{sec:thermalization}, such thermally produced Q-balls disappear by the time the circular rotation dominates the universe, and the analyses in the following sections are unaffected. In summary, production of Q-balls from a thermal potential is harmless, and Q-balls produced from the vacuum potential---that might give dangerous isocurvature perturbations---can be avoided by the choice of flat direction or the superpartner spectrum.

There is another process that can create inhomogeneities in $P$. Even for a logarithmic potential with a positive $K$, before thermalization when the rotation has non-zero ellipticity, fluctuations of $P$ around the rotating background grow~\cite{Co:2020dya,Co:2020jtv} via parametric resonance~\cite{Dolgov:1989us,Traschen:1990sw,Kofman:1994rk,Shtanov:1994ce,Kofman:1997yn}. If the thermalization occurs after parametric resonance is efficient, the amplitude of the fluctuations of $P$  becomes as large as the zero mode amplitude.

Parametric resonance, however, does not affect production of the curvature perturbations discussed in Sec.~\ref{sec:perturbation}. Even if large fluctuations are produced by parametric resonance, they are limited to sub-horizon scales since the super-horizon modes do not have instability.  This means parametric resonance leaves unaffected---on the super-horizon scales---both the $U(1)$ charge density $n_\theta$ and fluctuations imprinted in it during inflation.  Moreover, once thermalization occurs, $P$ becomes homogeneous within the Hubble horizon, and since the horizon size at this time is much larger than the inverse of the frequency of the rotation $\dot{\theta}^{-1}$, the gradient term is negligible in the dynamics of the rotation. The energy density of the rotation evolves as in the coherent case. Nevertheless, as we will see in Sec.~\ref{sec:fluctuations}, parametric resonance plays a crucial role in avoiding isocurvature and domain wall problems when the $U(1)$ symmetry is the PQ symmetry.

\subsection{Entropy production from the rotation}
\label{sec:washout}

In order for the rotation to act as a curvaton, the energy density of the rotation must be transferred into the bath. Indeed, the $U(1)$ charge in the rotation can be slowly washed out via explicit breaking of the $U(1)$ symmetry and produce entropy. For washout to occur, all (approximate) symmetries that protect the charge must be explicitly broken.

To illustrate this, consider a toy model of QCD with two flavors $u$ and $d$ where the $U(1)$ symmetry has a QCD anomaly. The Boltzmann equations governing the charge density are
\begin{align}
    \dot{n}_u = - \gamma_u n_u - \gamma_{\rm ss} \left(n_u + n_d - \frac{1}{2}\dot{\theta} T^2 \right) \nonumber \\
    \dot{n}_d = - \gamma_d n_d - \gamma_{\rm ss}\left(n_u + n_d - \frac{1}{2}\dot{\theta} T^2\right) \nonumber \\
    \dot{n}_\theta = - \gamma_{\rm ss}\left(n_u + n_d - \frac{1}{2}\dot{\theta} T^2\right).
    \end{align}
Here $n_u$ and $n_d$ are chiral asymmetry of $u$ and $d$, respectively, and $\gamma_{u,d}$ and $\gamma_{\rm ss}$ are the chiral symmetry breaking rates arising from the masses of $u,d$ and the strong sphaleron processes,
\begin{align}
    \gamma_{u,d} \simeq \alpha_3 m_{u,d}^2/ T, ~~
    \gamma_{\rm ss} \simeq 100 \alpha_3^5 T.
\end{align}
Solving for $n_{u,d}$ by taking $\dot{n}_{u,d} =0$ and substituting them into the equation for  $\dot{n}_\theta$, we obtain
\begin{align}
    \dot{n}_\theta = -\left(\gamma_u^{-1}+\gamma_d^{-1} + \gamma_{\rm ss}^{-1}\right)^{-1}\frac{T^2}{2S^2}n_\theta \equiv -\gamma_{\theta} n_\theta,
\end{align}
where we have used $n_{\theta} = -S^2 \dot{\theta}$. One can see that the washout rate for $n_{\theta}$ vanishes if the up- or down-quark mass is zero despite the explicit $U(1)$ symmetry breaking by the QCD anomaly. This is because a linear combination of the $U(1)$ symmetry and a quark chiral symmetry remains exact in the massless quark limit. Hence, the washout rate of the $U(1)$ charge $\gamma_\theta$ is roughly given by
\begin{align}
    \gamma_{\theta} \simeq {\rm min}(\gamma_u,\gamma_d,\gamma_{\rm ss}) \times \frac{T^2}{2S^2}.
\end{align}
That is, the washout rate is given by the rate of the bottleneck process times a further suppression factor $T^2/(2S^2)$. This additional suppression comes from the smallness of the chiral asymmetries $n_{u,d}$ at the quasi-equilibrium state, $n_{u,d}\sim \dot{\theta} T^2 \sim n_\theta T^2/(2S^2)$. 

Going beyond our toy model, washout generically requires  all linear combinations of the $U(1)$ symmetry with other symmetries be explicitly broken. As an example, consider the $\bar{u}_1\bar{u}_2\bar{d}_2\bar{e}_1$ supersymmetric flat direction with $S > T $, which completely breaks $SU(3)_c\times U(1)_Y$. This breaking means that the strong sphaleron rate is exponentially suppressed, so that the chiral symmetries of $\bar{u}_1$, $\bar{u}_2$, and $\bar{d}_1$ are broken only by $y_u$, $y_c$, and $y_d$, respectively. The $\bar{e}_1$ chiral symmetry is broken by $y_e$. (The chiral symmetry breaking rates are given by $y^2S^2/T$ rather than by $y^2 T$, assuming $yS <T$.) $B+L$ is broken by the weak sphaleron process. $L_1-L_2$ and $L_1-L_3$ can be broken by slepton mixing. $B_1-B_2$ and $B_1-B_3$ are broken by the CKM mixing and/or squark mixing. The PQ (an approximate symmetry in the MSSM in the limit of $\mu =0$) and $R$ symmetry must be also broken, since otherwise a linear combination of these symmetries and $B+L$ remains unbroken. The PQ symmetry is broken by the $\mu$ or $B\mu$ term, while the $R$ symmetry is broken by the gaugino mass, the $B\mu$ term, or $A$ terms. The washout rate will be controlled by the minimum of all these breakings. A similar story holds for $QQQL$ directions, but in this case, since the $SU(2)_L$ gauge symmetry is broken by the flat direction, effective $B+L$ violation by weak sphalerons can occur only for small $S$; see the discussion below for the modification of the washout rate for this case.

Some of the MSSM flat directions have non-zero $B-L$ charges. For this case, the washout of the $B-L$ charge can be achieved by $R$-parity violation or light right-handed neutrinos. The 
curvaton could also be
a flat direction with neither $B$ nor $L$ charge but with flavor charges, such as $Q_1 \bar{u}_2 L_1 \bar{e}_2$. Washout of the flavor charges can occur via $y_c$, $y_\mu$, $1\mathchar`-2$ squark or CKM mixing, and $1\mathchar`-2$ slepton mixing. This flat direction also has non-vanishing PQ and $R$ charge, so the $\mu$ or $B\mu$ term and a source of $R$-symmetry breaking are also required.

In all cases, the washout may be initially ineffective because of the rate suppression at large $S$, and the rotation is stable. However, as $S$ decreases the washout becomes effective and the entropy production from the rotation begins. The temperature of the universe during the entropy production is given by
\begin{align}
\label{eq:T}
    T^4 \sim \rho_\theta \frac{\gamma_\theta}{H}.
\end{align}
Since this equation depends on $T$, $\rho(n_\theta)$, $S(n_\theta)$, and $H(\rho_\theta(n_\theta),T)$, the temperature $T$ after the entropy production begins is determined as a function of $n_\theta$. The uniform energy density slice is therefore the uniform $n_\theta$ slice. As we will see in Sec.~\ref{sec:perturbation}, this enables the computation of the curvature perturbation via the $\delta N$ formalism~\cite{Sasaki:1995aw,Wands:2000dp,Lyth:2004gb} using the uniform $n_\theta$ slice. The rotation is completely washed out once $\gamma_\theta$ becomes larger than $H$.

The above estimation of the washout rate implicitly assumes $m_S<T$. As discussed in Sec.~\ref{sec:evolution}, the rotation provides chemical potentials to particles in the thermal bath $\mu_{\rm bath} \sim \dot{\theta} = m_S$.
So for $m_S < T$, the distribution of particles in the thermal bath is nearly identical to one with $\mu_{\rm bath} =0$, and the energy density of the thermal bath as a whole and the typical energy of particles are not affected by the presence of the rotation. However, if washout does not occur before $T$ drops below $m_S$, the energy density of the bath is affected by the presence of the large chemical potential. Indeed, the energy density of the bath with $T \ll \mu_{\rm bath}$ is kept at $\rho_{\rm bath} \sim \mu_{\rm bath}^{4} \sim m_S^4$ and the typical energy of particles in the thermal bath at $\mathcal{O}(m_S)$. This deviation from the standard scaling of the energy density of radiation $\propto R^{-4}$ is due to significant energy transfer from the rotation to the bath.  We emphasize that this occurs even for $U(1)$ charge-conserving (not washout) processes. The radiation energy of the universe dominantly comes from the rotation, and the uniform-$n_\theta$ slice is again a uniform energy density slice. The $U(1)$ charge should ultimately be washed out, so that the universe eventually reverts to a standard radiation-dominated universe. To find the washout rate in this phase, we may replace $T$ with $m_S$ in our earlier estimate of the washout rate. In this phase, prior to washout, the temperature drops in proportion to $R^{-3}$, as can be seen from the entropy density $\propto \mu_{\rm bath}^2 T \sim m_S^2 T$, so $T$ becomes much smaller than $m_S$.%
\footnote{The decrease of the temperature can be slower because of the entropy production from the partial washout of the rotation.}

If the washout rate is sufficiently small, $S$ may drop below $m_S (>T)$ before the $U(1)$ charge is washed out. For example, the washout of $QQQL$ involving non-perturbative processes of the weak interaction is ineffective for $S> \mu_{\rm bath} \sim m_S$~\cite{Rubakov:1985ehm,Rubakov:1985nk}, and so this stage is reached.
In this stage, the chemical potential of the system is no longer kept at $\mathcal{O}(m_S)$; since the $U(1)$ charge in the rotation of $P$ does not dominate the total charge, a fixed $\mu_{\rm bath} \sim m_S$ would mean that the total $U(1)$ charge would not decrease in proportion to $R^{-3}$, which is in contradiction with charge conservation. Thus, the chemical potential must decrease. Since the temperature and the chemical potential become smaller than $m_S$, the abundance of $P$ (and hence the field value of $P$) is exponentially suppressed, and the $U(1)$ charge is mostly stored in the particle-antiparticle asymmetry of particles lighter than $m_S$, such as SM particles. The temperature as well as the chemical potential decrease in proportion to $R^{-1}$. 
The universe is dominated by radiation. The factor $T^2/S^2$ in the washout rate is absent. At this point, if $P$ is an MSSM flat direction, since there is no suppression of the washout rate by a large field value of $S$, the washout of the $U(1)$ charge is effective and the chemical potentials will eventually vanish.%
\footnote{Exceptions are lepton flavor charges $L_i-L_j$, whose washout may be ineffective if the slepton mass matrices are very nearly diagonal. See~\cite{Co:2021qgl} for the estimation of the washout rate involving sfermion mixing.  This would result in $Y_{L_i-L_j} \sim 1/g_* \sim 0.01$, until they are washed out by the neutrino oscillation at 1-10 MeV~\cite{Dolgov:2002ab,Wong:2002fa}. The large lepton flavor charge leads to non-zero baryon asymmetry even if $B-L=0$~\cite{Mukaida:2021sgv}.}
Note that once $P$ relaxes to a smaller field value, the weak gauge symmetry is restored, and the non-perturbative $B+L$-violating processes no longer receive an exponential suppression.

%%%%%%%%%%%%%%%%%%%%%%%%%%%%
\section{Cosmic perturbations and non-Gaussianities}
\label{sec:perturbation}
%%%%%%%%%%%%%%%%%%%%%%%%%%%%

During inflation, the angular direction $\theta$ remains nearly massless. As is the case for any massless field~\cite{Mukhanov:1981xt,Hawking:1982cz,Starobinsky:1982ee,Guth:1982ec,Bardeen:1983qw}, inflation imprints fluctuations along the $\theta$ direction~\cite{Steinhardt:1983ia,Linde:1985yf,Seckel:1985tj}, and the fluctuation $\delta\theta({\bf x})$ follows a Gaussian distribution with a width $H_{\rm inf}/(2\pi S_{\rm inf})$. The fluctuation in the angular direction leads to the fluctuation of the magnitude of the kick discussed in the previous section.  This, in turn, leads to variations in the charge density $n_\theta$. In this section, we discuss how the curvature perturbations of the universe arise from the fluctuation of $n_\theta$.

\subsection{Curvature perturbations}
We compute the spectrum of the curvature perturbations with the $\delta N$ formalism~\cite{Sasaki:1995aw,Wands:2000dp,Lyth:2004gb}, which relies on a computation of the number of e-foldings, $N$, between an initial flat time slice and a final uniform-density slice. The fluctuations in the number of e-foldings between these slices $\delta N(\bold{x})$ give the curvature perturbations $\zeta(\bold{x})$.

We take the initial flat time slice $t_i$ at a time where the energy is dominated by the inflaton (or the radiation produced by it) and the $U(1)$ charge is conserved. In this slicing, the $U(1)$ charge has spatial fluctuations that originate from the fluctuation of the initial value of the angular direction produced during inflation; recall the $U(1)$ charge is created by the kick to the angular direction whose magnitude depends on the angle, see Eqs.~(\ref{eq:potential}) and (\ref{eq:nthetaeom}). Fluctuations that are observable in the CMB and large scale structure formation correspond to long-wavelength modes, and so the evolution of the charge density is unaffected by gradients in the energy density. Spatial fluctuations are therefore completely determined by the initial value of the angular direction $\theta_i$ before the initiation of the rotation, so we can parameterize the charge density by
\begin{align}
\label{eq:ni}
    n_\theta(t_i,{\bf x}) \equiv n_{i}(\theta_i({\bf x})).
\end{align}
Since the radiation density of the universe is eventually dominated by that created from the (partial) washout of the rotation, see  Eq.~(\ref{eq:T}), a slice with uniform charge will give a slice with uniform energy density. We therefore take the final slice $t_f$ to be a uniform charge slice,
\begin{align}
    n_\theta(t_f,{\bf x}) = n_\theta(t_f).
\end{align}
The number of e-foldings between the two slices, $N$, is given by
\begin{align}
    n_\theta(t_i,{\bf x}) \times {\rm exp}( - 3 N(t_i,t_f,{\bf x})) = n_\theta(t_f).
\end{align}
Curvature perturbations are given by the variation of $N$ between the two time slices~\cite{Sasaki:1995aw,Wands:2000dp,Lyth:2004gb}:
\begin{align}
\label{eq:zeta}
    \zeta({\bf x}) = \delta N({\bf x}) =  \frac{1}{3} \delta {\rm ln}(n_{i}(\theta_i({\bf x}))) = \frac{1}{3} \frac{n_{i}^\prime}{n_{i}}\delta \theta({\bf x}) + \frac{n_{i} n_{i}^{\prime\prime} - n_{i}^{\prime2}}{6 n_{i}^2} (\delta \theta({\bf x}))^2 + \cdots,
\end{align}
where the primes denote the derivative with respect to $\theta_i$. From the fluctuation $\delta \theta \sim H_{\rm inf}/(2\pi S_{\rm inf})$,
we obtain a power spectrum of the curvature perturbations
\begin{align}
\label{eq:Pzeta}
   P_\zeta (k) = \left(\frac{n_{i}^\prime}{3n_{i}}\right)^2 \left(\frac{H_{\rm inf}(k)}{2\pi S_{\rm inf}(k)}\right)^2.
\end{align}
The observed value is $P_{\zeta}(k) = 2.1 \times 10^{-9}$ at  $k= 0.05$ (Mpc)$^{-1}$~\cite{Planck:2018jri}, so the  inflation scale may be related to the value of $S$ during inflation $S_{\rm inf}$ and the charge density as
\begin{align}
    H_{\rm inf} \simeq 9 \times 10^{11} \GeV 
    \left( \frac{S_{\rm inf}}{10^{16}\GeV} \right) \left(\frac{10}{(n_i^\prime/3n_i)^2}\right)^{ \scalebox{1.01}{$\frac{1}{2}$} }.
\end{align}

Starting with the potential in Eq.~(\ref{eq:potential}), the dependence of the function $n_{i}$ on $\theta_i$ can be evaluated analytically in the limit where the kick in the angular direction is weak ($a_{\cancel{U(1)}} \ll 1$ or a $\theta_{i}$ close to an integer multiple of $\pi/(n+1)$) and the ellipticity of the rotation is large. In this case, the right-hand side of Eq.~(\ref{eq:nthetaeom}) can be treated perturbatively, neglecting the change of the angle around the time of the initiation. We then find
\begin{align}
\label{eq:f_analytic}
     \lim_{a_{\cancel{U(1)}} \rightarrow 0} n_i(\theta_i) \propto \frac{\partial V}{\partial \theta_i} \propto {\rm sin}((n+1) \theta_i).
\end{align}
In this limit,
\begin{align}
\label{eq:zetafactor}
    \lim_{a_{\cancel{U(1)}} \rightarrow 0} \left(\frac{n_{i}^\prime}{3n_i}\right)^2 = (n+1)^2\frac{\cos^2{((n+1)\theta_i)}}{9  \sin^2{((n+1)\theta_i)}}.
\end{align}
Note the enhancement of the fluctuation for large $n$.

\begin{figure}
\includegraphics[width=0.9\linewidth]{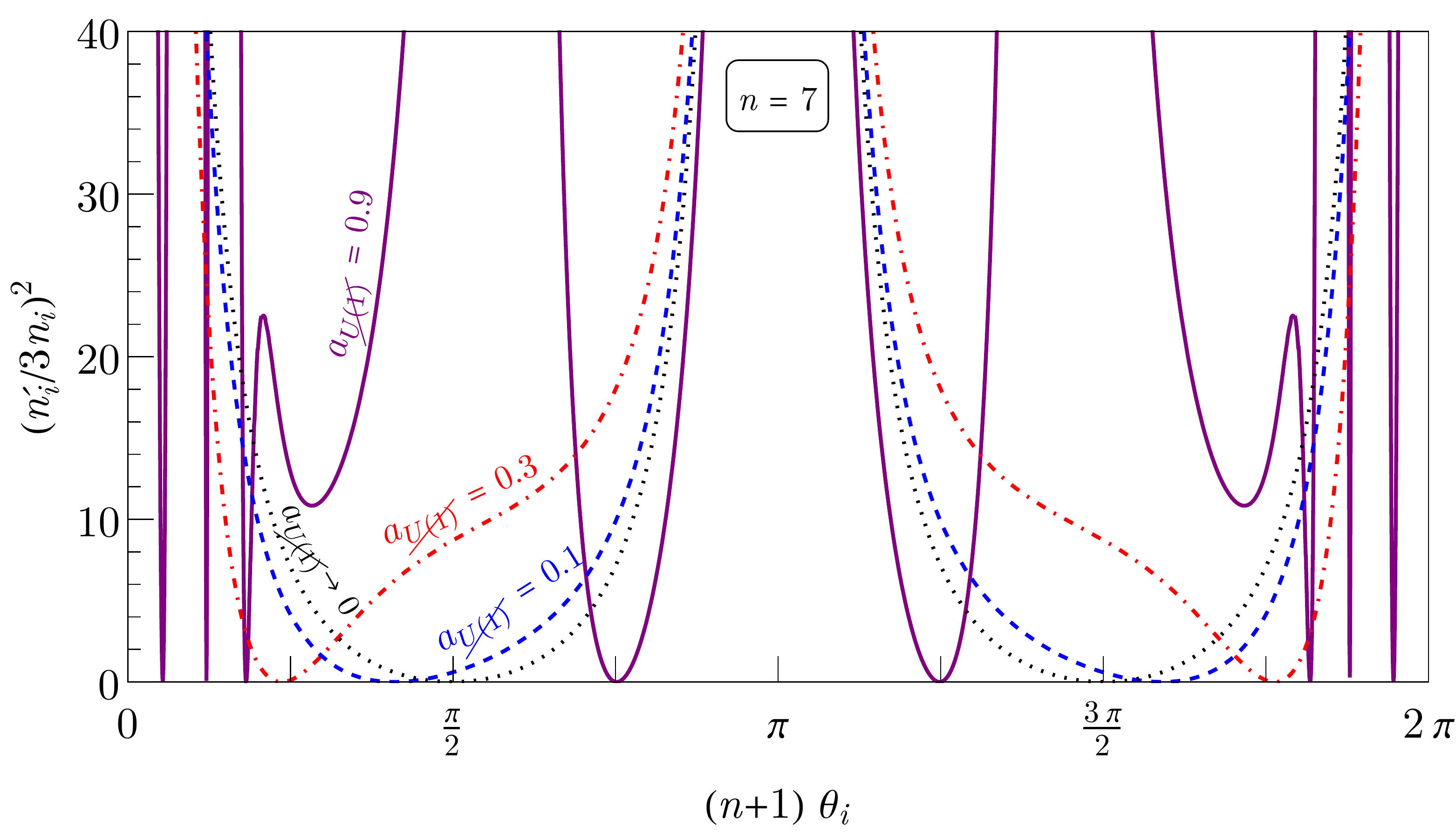}
\caption{The dependence of $P_\zeta / (H_{\rm inf}/2\pi S_{\rm inf})^2 = (n_{i}^\prime/3n_i)^2$ on the initial angle $\theta_i$ for $n=7$ and several values of the $U(1)$ breaking parameter $a_{\cancel{U(1)}}$ defined in Eq.~(\ref{eq:potential}). Here $n_i$ is a charge density in an initial flat time slice defined in Eq.~(\ref{eq:ni}).}
\label{fig:zeta}
\end{figure}

If the kick is strong and the rotation becomes nearly circular, analytical evaluation is difficult. In this regime, we employ a numerical computation as described in Appendix~\ref{sec:charge}. The resultant value of $(n_{i}^\prime/3n_{i})^2$ is shown in Fig.~\ref{fig:zeta} for several reference points. As the kick becomes weak, the result approaches the analytical approximation, while for $a_{\cancel{U(1)}} = {\mathcal O}(1)$, the deviation becomes significant.
When $(n+1)\theta_i/\pi$ is close to an integer, the kick is weak for all values of $a_{\cancel{U(1)}}$ and the analytical approximation works well. Other non-trivial features of the curves can be understood from $n_i(\theta_i)$ in Fig.~\ref{fig:charge}, which is described in Appendix~\ref{sec:charge}.
In particular, for $a_{\cancel{U(1)}}$ close to unity, the angular direction oscillates before the rotation begins and $n_i(\theta_i)$ non-trivially depends on $\theta_i$.
In these figures, it is assumed that the universe is radiation dominated when the rotation is initiated, but the case with matter domination is qualitatively similar. Here we take $n=7$, which corresponds to the MSSM flat directions lifted by $W = (\bar{u}\bar{u}\bar{d}\bar{e})^2$, $(QQQL)^2$, $(Q\bar{u} Q \bar{d})^2$, $(Q \bar{u} L \bar{e})^2$~\cite{Gherghetta:1995dv}.  For larger $n$ with a fixed $a_{\cancel{U(1)}}$, the kick becomes stronger (see Eq.~(\ref{eq:newvars})), so the deviation from Eq.~(\ref{eq:zetafactor}) becomes larger.

%%%
%%%

We now comment on the case with Q-ball formation discussed in Sec.~\ref{sec:Qball_PR} and the impact of such Q-balls on perturbations. When the potential of $S$ is flatter than a quadratic one, fluctuations around the rotation have instability modes. Q-balls form when the growth rate of the fluctuations by the instability becomes comparable to the Hubble expansion rate, which means that the total energy density of the universe when this condition holds is fixed.%
\footnote{If the potential of $S$ is significantly is different from the quadratic one (e.g., the thermal log potential and gauge-mediated contributions above the mediation scale), the growth rate can depend on $n_\theta$. Still, when the growth rate becomes comparable to the Hubble expansion rate is not uniquely determined by $n_\theta$, so the modulated reheating contribution discussed below persists.}
If the production occurs before the circular rotation comes to dominate,  the uniform-density slice does not coincide with the uniform-$n_\theta$ slice, so $n_\theta$ fluctuates when the Q-balls form. The charge of Q-balls depends on $n_\theta$ when they are produced~\cite{Kusenko:1997si,Hiramatsu:2010dx,Kasuya:2000wx,Doddato:2011fz} so also fluctuates.
The lifetime of the Q-balls, which is regulated by the Pauli-blocking near their surface~\cite{Cohen:1986ct}, depends on the charge, so the lifetime of the Q-balls also fluctuates. When the Q-balls evaporate, this induces extra curvature perturbations~\cite{Harigaya:2019uhf}, as in the modulated reheating scenario~\cite{Dvali:2003em,Kofman:2003nx}.%
\footnote{If the microscopic decay rate of the Q-ball constituents is much smaller than the evaporation rate (not the case for MSSM flat directions in gravity mediation), the Q-ball decay rate is determined by the microscopic one, and the modulated reheating effect is absent. Still, if the $S$ potential significantly deviates from a quadratic one, as in gauge mediation, the energy per charge can depend on the charge of Q-balls~\cite{Dvali:1997qv}; this may create extra curvature perturbations.}
Although there is no problem in this scenario as far as the curvature perturbations are concerned, if the baryon asymmetry is also produced from the rotation, this extra source of curvature perturbations leads to too-large correlated baryon isocurvature perturbations. As we discussed in Sec.~\ref{sec:Qball_PR}, the vacuum potential does not support the Q-ball solution for a class of MSSM flat directions and PQ symmetry breaking fields.

However, even if the vacuum potential does not allow for the Q-ball solution, the thermal potential may. As we discuss in Appendix~\ref{sec:thermalization}, however, such thermally induced Q-balls disappear by the time the rotation would dominate the universe. After the Q-balls disappear, the rotation becomes nearly homogeneous again, and the above analysis is applicable. Therefore, vacuum potentials that do not admit Q-ball formation, such as those identified in Sec.~\ref{sec:Qball_PR}, will be free of the isocurvature problem discussed above.

\subsection{Spectral index}

The spectral index in the rotating case is given by
\begin{align}
    n_s = 1- 2\epsilon \left(\frac{n-2}{n-1}\right)  + 2 \eta_{\theta\theta},~~ \eta_{\theta\theta} = \frac{V_{\theta\theta}}{S_{\rm inf}^2H_{\rm inf}^2},
\end{align}
where $\epsilon$ is the first slow-roll parameter of the inflaton. This is similar to the usual curvaton scenario, but with an extra factor $(n-2)/(n-1)$, which comes from the dependence of $S_{\rm inf}$ on $H_{\rm inf}$.  The observed spectral index $n_s = 0.9665 \pm 0.0038$~\cite{Planck:2018jri} requires that $\epsilon \simeq 0.02$ or $\eta_{\theta\theta} \simeq - 0.02$. The former requires large field inflation models~\cite{Lyth:1996im}. The latter would require explicit $U(1)$ breaking. This cannot originate from the $a_{\cancel{U(1)}}$ term in Eq.~(\ref{eq:potential}); the mass of the angular direction there is comparable to the Hubble scale at the time of the kick ($3H \simeq m_S$) and is much smaller than the Hubble scale during inflation.%
\footnote{While this is not the case if the Hubble scale during inflation is not much above $m_S$, to explain the magnitude of the curvature perturbations, the required $S_{\rm inf}$ is small (see Eq.~\eqref{eq:Pzeta}). In this case the rotation cannot dominate the universe unless $m_S$ is above $10^9$ GeV.}
An extra $U(1)$-breaking is necessary.

An extra $U(1)$-breaking potential may be provided by an extra $U(1)$-breaking term in either the super- or K\"ahler-potential. The resultant $U(1)$-breaking potential must be of higher power in $P$ than that in Eq.~(\ref{eq:potential}), so that the $U(1)$ breaking is negligible by the time $P$ begins rotation while it can give $\eta_{\theta\theta}\sim -0.01$ during inflation. 
 
Alternately, explicit $U(1)$ breaking may be provided from an extra $A$ term associated with the superpotential term in Eq.~(\ref{eq:PQV}) from coupling to the inflaton sector.
Such a term, namely, a Hubble-induced $A$ term, readily arises in inflation models without an $R$ symmetry. Even for $R$-symmetric models, a Hubble-induced $A$ term can arise from spontaneous $R$ symmetry breaking in the inflaton sector and the K\"ahler potential
\begin{align}
    K = \frac{1}{M_*^2} ZZ^\dag PP^\dag,
\end{align}
where $Z$ is a chiral multiplet whose $F$ term ($= \sqrt{3} H_{\rm inf} \MPl$) is responsible for the inflaton potential with $\MPl$ the reduced Planck mass, and $M_*$ is the cut-off scale. The resultant $A$ term potential is as large as 
\begin{align}
    n Z_{\rm inf}\frac{H_{\rm inf}\MPl }{M_*^2} \frac{P^{n+1}}{M^{n-2}} + {\rm h.c.}
\end{align}
The axion mass from this $A$ term is $\mathcal{O}(0.1) H_{\rm inf}$ if $n^{3/2}Z_{\rm inf}\MPl/M_*^2$ is $\mathcal{O}(0.01)$. Here we use Eq.~(\ref{eq:Pinf}) to evaluate the field value of $P$. For example, for $M_* \sim \MPl$ and $n=10$, the required $Z_{\rm inf}$ is $\mathcal{O}(0.001) \MPl$ and is much below the Planck scale. Because of the smallness, the extra $A$ term of the required size can be easily obtained even in small field inflation models such as new inflation and hybrid inflation. The required $Z_{\rm inf}$ is even smaller if $M_*$ is below the Planck scale. Note that, however, the Hubble induced $A$ term can increase during inflation if $Z_{\rm inf}$ increases during inflation. If the axion mass from the Hubble induced $A$-term becomes as large as the Hubble scale, the fluctuation of the angular direction is damped and the curvature perturbation cannot be explained. Also, if the $R$ symmetry is (approximately) continuous, a linear combination of the phase degree of freedom of the inflaton and the axion remains massless. In order to obtain $\eta_{\theta\theta}\sim - 0.01$, the massless direction must be dominantly the inflaton component. We leave the investigation of the compatibility of our scenario with various inflation models for future work.

\subsection{Non-Gaussianity}
Just like the case of the oscillating curvaton, the curvature perturbations created from the axion rotation have non-Gaussianity of the local type. The local non-Gaussianity parameter $f_{\rm NL}$ is defined by
\begin{align}
    \zeta({\bf x}) \simeq g({\bf x}) + \frac{3}{5}f_{\rm NL} \times g({\bf x})^2,
\end{align}
where $g({\bf x})$ follows a gaussian distribution. Comparing this with Eq.~(\ref{eq:zeta}), we obtain
\begin{align}
    f_{\rm NL} = -\frac{5}{2} \left(1 - \frac{n_{i} n_{i}^{\prime\prime}}{n_{i}^{\prime 2}}\right).
\end{align}
The non-Gaussianity becomes large as $n_i'$ approaches $0$, since small $n'$ means that the first order,Gaussian part of $\zeta$ becomes small and the second order part that leads to non-Gaussianity is relatively enhanced. This is in contrast to the curvaton from an oscillating field with a mass term, where $f_{\rm NL}= -5/4$ when the curvaton dominates, and larger $|f_{\rm NL}|$ requires that the curvaton is subdominant when it decays. 

\begin{figure}
\includegraphics[width=0.9\linewidth]{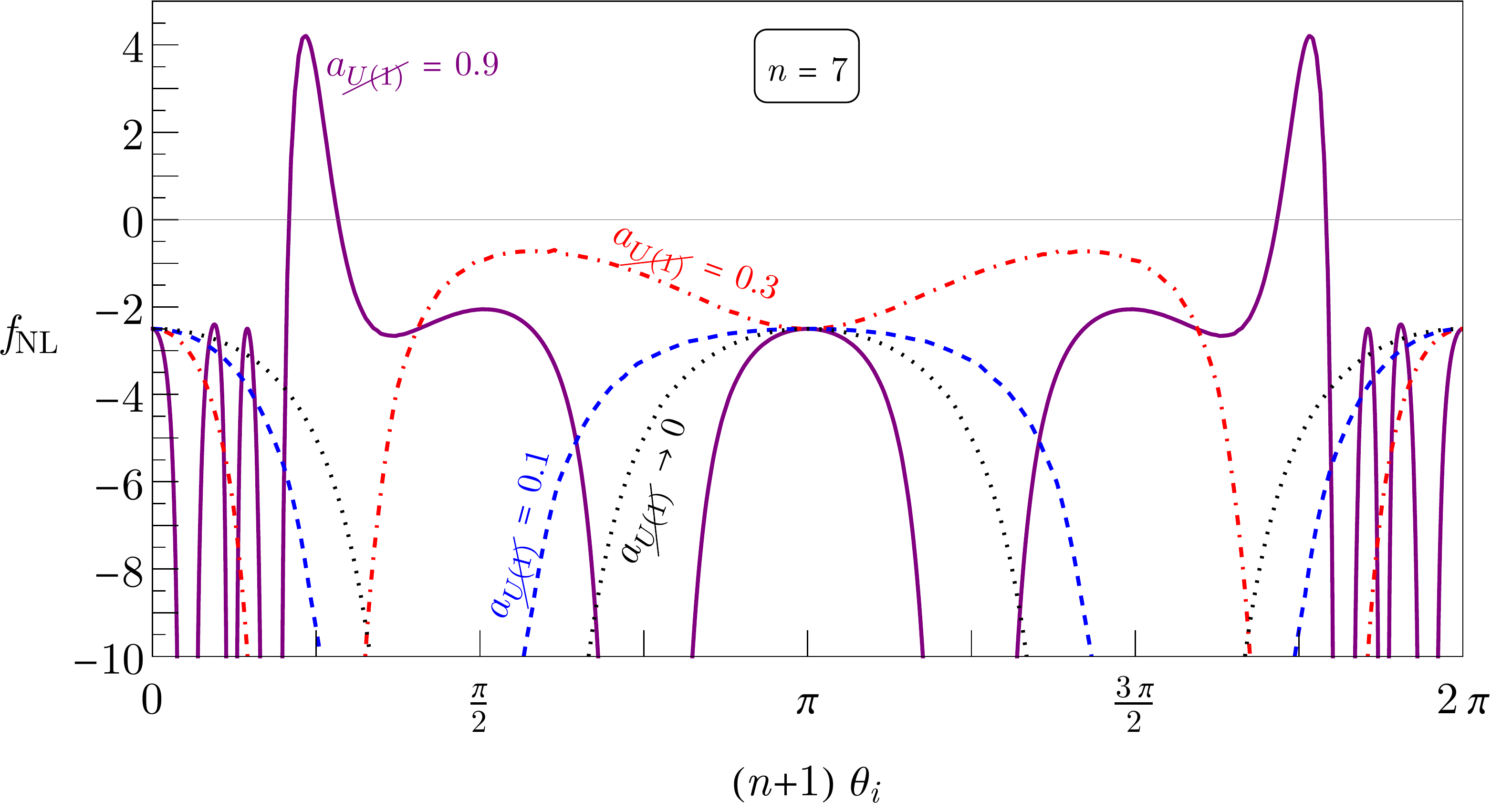}
\caption{The dependence of the non-Gaussianity parameter $f_{\rm NL}$ on the initial angle $\theta_i$ for $n=7$ and several values of the $U(1)$-breaking parameter $a_{\cancel{U(1)}}$ defined in Eq.~(\ref{eq:potential}).}
\label{fig:fNL}
\end{figure}

In the limit of weak kick, $f_{\rm NL}$ can be analytically obtained using Eq.~(\ref{eq:f_analytic}),
\begin{align}
\label{eq:fNLanalytic}
    \lim_{a_{\cancel{U(1)}} \rightarrow 0} f_{\rm NL} = - \frac{5}{2  \cos^2{((n+1)\theta_i)}} \le -2.5 .
\end{align}
This is consistent with the bound from the observations of the CMB, $f_{\rm NL} = -0.9 \pm 5.1~(68\%~{\rm C.L.})$~\cite{Planck:2019kim} for a majority of $\theta_i$, and can be detected via future observations of small scale structures~\cite{Alvarez:2014vva,Dore:2014cca} that are sensitive to $\mathcal{O}(1)$ $f_{\rm NL}$. A numerical evaluation of $f_{\rm NL}$ for several reference points are given in Fig.~\ref{fig:fNL}. When the kick is strong, it largely deviates from the elliptical case, but the magnitude of $f_{\rm NL}$ is generically ${\mathcal O}(1)$ or larger, although it can be much smaller for a fine-tuned $\theta_i$ for $a_{\cancel{U(1)}} \sim 1$. Note that the sign of $f_{\rm NL}$ can be positive for $a_{\cancel{U(1)}} \sim 1$, but only for a small range of $\theta_i$. Here we take $n=7$. As mentioned before, a larger $n$ corresponds to a larger kick for fixed $a_{\cancel{U(1)}}$, which means that the deviation from the analytical estimation in Eq.~(\ref{eq:fNLanalytic}) becomes larger. The conclusion that a sizable $f_{\rm NL}$ is produced for all but perhaps a small range of $\theta_{i}$ is robust to the choice of $n$.

%%%
%%%

\section{Rotating axion field}
\label{sec:axion}

Having discussed general dynamics of rotating fields and how perturbations are induced in this scenario, we now discuss the case where a rotating axion acts as a curvaton. The explicit breaking in Eq.~(\ref{eq:PQV}) is expected in theories where the PQ symmetry arises as an accidental symmetry~\cite{Georgi:1981pu,Holman:1992us,Barr:1992qq,Kamionkowski:1992mf,Dine:1992vx}. Most of the discussion for generic complex fields is applicable to the PQ symmetry breaking field, but here we discuss few differences arising from the spontaneous breaking of the symmetry at the vacuum.

\subsection{Nearly quadratic potential and kination phase}

Thus far we have assumed a nearly quadratic potential at large $S$.  For a PQ breaking field, this may be explicitly realized in different ways. For example, the potential may be given by the supersymmetry breaking soft mass of $P$ that is positive at UV but becomes negative at IR by renormalization group running~\cite{Moxhay:1984am},
\begin{align}
\label{eq:dim_trans}
V(P) = m_S^2 |P|^2 \left( {\rm ln} \frac{2 |P|^2}{f_a^2} -1 \right).
\end{align}
Another example is a model with two PQ symmetry breaking fields $P$ and $\bar{P}$,
\begin{align}
\label{eq:two_field}
W = X( P \bar{P} - v_{P}^2 ),~~V_{\rm soft} = m_P^2 |P|^2 +  m_{\bar{P}}^2 |\bar{P}|^2,
\end{align}
where $X$ is a chiral field whose $F$-term fixes $P$ and $\bar{P}$ on the moduli space $P\bar{P} = v_{P}^2$. For $P \gg v_P$ or $\bar{P} \gg v_P$, the saxion potential is again nearly quadratic with $m_S \simeq m_P$ or $m_{\bar{P}}$, respectively. However, the latter model has difficulty in avoiding a domain-wall problem as we will see in Sec.~\ref{sec:fluctuations}.

In these models, Eq.~(\ref{eq:dtheta}) and $n_\theta \propto R^{-3}$ require that~\cite{Co:2019wyp}
\begin{align}
\begin{cases}
    \dot{\theta} \propto R^0 & \\
    S^2 \propto R^{-3} & :S \gg f_a, \\
    \rho_\theta \propto R^{-3}&
\end{cases}
~~~~~~
\begin{cases}
    \dot{\theta} \propto R^{-3} & \\
    S^2 \propto R^{0} & :S \simeq f_a. \\
    \rho_\theta \propto R^{-6}&
\end{cases}
\end{align}
Unlike the case where the $U(1)$ symmetry is not spontaneously broken at the vacuum, the rotation eventually evolves as kination $(\rho_\theta \propto R^{-6})$ and redshifts faster than radiation does. Therefore, the PQ charge does not have to be washed out completely---cosmic expansion will eventually make the axion rotation subdominant. To successfully act as a curvaton, it is enough that the entropy of the universe dominantly comes from the washout of the rotation.

The kination domination preceded by matter domination imprints a peculiar signal on the spectrum of primordial gravitational waves~\cite{Co:2021lkc,Gouttenoire:2021wzu,Gouttenoire:2021jhk} because it changes the expansion history of the universe relative to the standard radiation-dominated universe.  In our setup, because of the entropy production from the rotation, gravitational waves at high frequencies are suppressed in comparison with the spectrum shown in these references, leaving a different signature. Also, Refs.~\cite{Co:2021lkc,Gouttenoire:2021wzu,Gouttenoire:2021jhk} assumed that the PQ charge of the axion rotation decreases only by redshifting. As a result,
unless the kination domination ends at a sufficiently high temperature, the axion rotation overproduces axion dark matter by kinetic misalignment~\cite{Co:2019jts}. This puts a lower bound on the frequency of the gravitational waves that can be affected by the kination domination. If the axion rotation is mostly washed out, this lower bound can be relaxed or removed.

\subsection{Washout of PQ charge}

According to the generic discussion in Sec.~\ref{sec:washout}, the washout rate is given by the smallest of the strong sphaleron rate and chiral symmetry breaking rates times $T^2/S^2$. However, unlike MSSM flat directions, $S$ stops decreasing once it reaches $S =f_a$ and the washout of the rotation is not guaranteed. Indeed, if the up-quark chiral symmetry is broken only by the up Yukawa coupling $y_u$ as in the SM, scattering involving this coupling can provide the bottleneck process, and hence the washout rate is given by
\begin{align}
    \gamma_\theta \simeq \alpha_3 y_u^2 \frac{T^3}{S^2}.
\end{align}
Since $y_u$ is small and $S$ stops decreasing at $f_a$,
the axion rotation cannot produce entropy and therefore  cannot act as the curvaton. The chiral symmetry breaking rate may be, however, larger in theories beyond the SM. For example, squark mixing in the MSSM can lead to more effective washout, and consequently, entropy production from the rotation~\cite{Co:2021qgl}.

\subsection{Fluctuations of the axion field}
\label{sec:fluctuations}

Although the energy density of the rotation is homogeneous in the uniform-$n_\theta$ time slice as discussed in Sec.~\ref{sec:evolution}, the angle $\theta$ may not be so. If fluctuations in $\theta$ are too large, domain wall and isocurvature problems can arise.

We now calculate the fluctuations in $\theta$.  The change in the angle from an initial time to the final time $t_f$ is
\begin{align}
    \Delta \theta = \int^{t_f} \dot{\theta} {\rm d}t = \int_{n_\theta(t_f)} \frac{n_\theta}{S^2} \frac{{\rm d}n_\theta}{\dot{n}_\theta} = - \frac{1}{3} \int_{n_\theta(t_f)} \frac{{\rm d}n_\theta}{H S^2}.
\end{align}
In the epoch prior to washout,  $H$ is not uniquely determined by $n_\theta$.  That means that in this integration, $H$ fluctuates.  This leads to a fluctuation of $\theta$ produced per Hubble time of 
\begin{align}
    \delta \theta \sim  \frac{n_\theta}{S^2} \frac{\delta H}{H^2} = \frac{\dot{\theta}}{H} \frac{\delta H}{H} \sim \frac{\dot{\theta}}{H} \frac{\delta \rho_{\rm tot}}{\rho_{\rm tot}},
\end{align}
where $\rho_{\rm tot}$ is the total energy density of the universe and $\delta \rho_{\rm tot}$ is its fluctuation defined on the uniform-$n_\theta$ slice, for which $\delta \rho_{\rm tot} = \delta \rho_{\rm rad}$. Using entropy conservation $\delta s /s = {\rm const}$, one can see that $\delta \rho_{\rm tot}/\rho_{\rm tot} \propto \rho_{\rm rad}/\rho_{\rm tot}$. Therefore, before the rotation dominates, $\dot{\theta}/H \propto R^2$ and $\delta \rho_{\rm tot}/\rho_{\rm tot} \propto R^0$. After the rotation dominates but before behaving as kination, $\dot{\theta}/H \propto R^{3/2}$ and $\delta \rho_{\rm tot}/\rho_{\rm tot} \propto R^{-1}$. During kination domination, $\dot{\theta}/H \propto R^0$ and $\delta \rho_{\rm tot}/\rho_{\rm tot} \propto R^2$. The fluctuation produced per Hubble time is larger at later times.

What is the minimum possible value of $\delta \theta$?  This occurs when the entropy is produced from the axion rotation immediately following the time when rotation comes to dominate.%
\footnote{In realistic situations, the entropy production can begin only much later, since the washout of the rotation requires multiple explicit symmetry breaking, including the strong sphaleron process, as is discussed in Sec.~\ref{sec:washout}. Entropy production immediately following domination could occur in principle if the PQ symmetry is explicitly broken by interactions other than QCD but in a way that the extra explicit breaking is negligible in the present universe so as not to spoil the solution to the strong CP problem.}
This minimizes the amount of time where $H$ is not determined by $n_{\theta}$. For this case,
\begin{align}
\label{eq:deltatheta}
\frac{\dot{\theta}}{H} & \simeq \frac{\dot{\theta}}{\dot{\theta} S_{\rm dom}/\MPl} = \frac{\MPl}{S_{\rm dom}},~~ \frac{\delta H}{H} \simeq \frac{\delta n_\theta}{n_\theta} = 3 \zeta, \nonumber \\
\delta \theta & \simeq \frac{\MPl}{S_{\rm dom}} 3\zeta \sim \frac{\MPl}{S_{\rm dom}} \times 10^{-4},
\end{align}
where $S_{\rm dom}$ is the field value of $S$ when the rotation energy density dominates. One might think that this fluctuation could be reduced by choosing a very large $S_{\rm dom}$.  However, the circular axion rotation can dominate the universe only after the radial mode is thermalized, and for large values of $S$, the interaction of $P$ with the thermal bath is suppressed and thermalization does not occur.

Let us then estimate the maximal value of $S_{\rm dom}$. It is maximized when the initial rotation has an $\mathcal{O}(1)$ ellipticity, dominates the universe,
and is thermalized with the largest possible thermalization rate. For this case, the domination occurs immediately following thermalization, upon which the bath inherits its energy density from the radial mode. The assumption of $\mathcal{O}(1)$ ellipticity ensures that the angular mode has a comparable density. For thermalization via a Yukawa coupling $y$ of $P$ with a fermion, the thermalization rate is $\Gamma_{\rm th} \simeq 0.1 y^2 T$. Since $yS < T$ is required for the fermion to be in the thermal bath,  $\Gamma_{\rm th} < 0.1 T^3/S^2$. With this bound on the rate, using $\Gamma_{\rm th}= 3H$ and $\pi^2 g_* T^4/30 = m_S^2 S^2 $, we obtain
\begin{align}
    S_{\rm dom} < 3 \times 10^{12} \GeV\times  \left(\frac{m_S}{10^6\GeV}\right)^{1/3}.
\end{align}
Combining this result with Eq.~\eqref{eq:deltatheta} means $\delta \theta \gg 1$ unless $m_S \gtrsim 5 \times 10^{11}$ GeV. Even with the extremely optimistic assumptions we have made,   $m_{S}> f_a$ would be required to avoid $\delta \theta \gtrsim 1$, and the PQ symmetry cannot be spontaneously broken by the potentials in Eqs.~(\ref{eq:dim_trans}) and (\ref{eq:two_field}). Therefore, large $\delta \theta$ is unavoidable unless there exists a more efficient thermalization mechanism.

The large $\delta \theta \gg 1$ leads to the production of domain walls around the epoch of the QCD phase transition. Unlike the case with spontaneous PQ symmetry breaking after inflation, there are no PQ strings in the universe. Therefore, these domain walls cannot have boundaries made of strings, and hence cannot disappear by shrinking even if the domain wall number is unity. Note that $\delta \theta$ originates from the fluctuation of the PQ charge, so exists in long-wavelength modes. Therefore, the size of the domain walls ranges from the horizon size to the super-horizon size. Although domain walls annihilate as they enter the horizon, large domain walls continue to enter the horizon, and the universe is eventually dominated by domain walls.

In addition to the domain wall problem, which results when $\delta \theta \gtrsim 1$, fluctuations in $\theta$ can lead to fluctuations in the dark matter abundance produced by the misalignment and kinetic misalignment mechanisms.  This will produce correlated dark matter isocurvature perturbations. This gives an even stronger constraint on $\delta \theta$.  The constraint can be analytically derived if the rotation stops before the axion can begin oscillation, $m_a \simeq 3H$. In this case, the axion abundance is simply determined by the misalignment contribution $\Omega_{\rm mis}h^2$~\cite{Preskill:1982cy, Dine:1982ah,Abbott:1982af} with the misalignment angle $\theta_{\rm mis}$ and the fluctuation $\delta \theta$ around it. Since $\Omega_{\rm mis}h^2\propto \theta_{\rm mis}^2$, the dark matter isocurvature perturbation is
\begin{align}
    {\cal P}_{S_{\rm DM}} = \frac{4 (\delta \theta)^2}{\theta_{\rm mis}^2} \left(\frac{\Omega_{\rm mis}h^2}{\Omega_{\rm DM}h^2}\right)^2 \simeq 10^{-6} (\delta \theta)^2 \theta_{\rm mis}^2 \left(\frac{f_a}{10^9\GeV}\right)^{2.38}.
\end{align}
The constraint from the CMB measurements ${\cal P}_{S_{\rm DM}}/{\cal P}_\zeta < 10^{-3}$~\cite{Planck:2018jri} gives
\begin{align}
    \delta \theta < 0.02 \frac{1}{\theta_{\rm mis}} \left(\frac{10^8 \GeV}{f_a}\right)^{1.19} .
\end{align}
Even for the smallest possible $f_a \sim 10^8$ GeV, for the natural value $\theta_{\rm mis}= \mathcal{O}(1)$, $S_{\rm dom} \gtrsim 10^{16}$ GeV is required; see Eq.~\eqref{eq:deltatheta}.  The thermalization must somehow occur right after the initiation of the rotation. Even the tuned case $\theta_{\rm mis} \sim \delta \theta$ requires $S_{\rm dom} \gtrsim 10^{15}$ GeV. If the rotation continues after the axion would begin oscillation, the axion dark matter is dominantly given by the kinetic misalignment mechanism. In this case, the axion dark matter abundance may non-trivially depend on $\delta \theta$. We do not discuss this case in this paper, but we expect a strong constraint from isocurvature perturbations.

These problems can be avoided by the restoration of the PQ symmetry. Here we discuss three ways in which the restoration of the PQ symmetry may occur---via thermal trapping, parametric resonance, or Q-ball formation---and how the domain wall and isocurvature problems from $\delta \theta$ are solved. The thermal trapping can occur after washout, so the curvaton has already imprinted adiabatic perturbations on the bath, and the trapping has no effect on their spectrum.  Parametric resonance and Q-ball formation occur before adiabatic perturbations are produced, but as discussed in Sec.~\ref{sec:Qball_PR}, they also do not affect the generation of adiabatic perturbations. 

Symmetry restoration via a thermal potential may occur for the one-field model of Eq.~(\ref{eq:dim_trans}). Let us derive the condition for this PQ restoration to occur. If the rotation is completely washed out when the field value of the radial direction is $S_{\rm wo}>f_a$, the temperature of the universe at that time is given by $m_S^2 S_{\rm wo}^2 = \pi^2 g_* T_{\rm wo}^4/30$. A Yukawa coupling $y$ of $P$ with fermions gives a thermal mass $\sim y T_{\rm wo}$ as long as fermions are in the bath  ($y f_a < T_{\rm wo}$). If this thermal mass exceeds the vacuum value, $y T_{\rm wo} > m_S$, $P$ will be trapped at the origin and $\delta \theta$ disappears. This removes both the domain wall and isocurvature problems. $y T_{\rm wo} > m_S$ and $y f_a < T_{\rm wo}$ may be simultaneously satisfied if $S_{\rm wo} \gtrsim 10 f_a$.  Even if $y f_a > T_{\rm wo}$, the two-loop thermal log correction~\cite{Anisimov:2000wx} gives a thermal mass $\sim \alpha_3 T^2/f_a$. This is larger than $m_S$ and the trapping at the origin occurs if $S_{\rm wo} \gtrsim 100 f_a$.  As discussed in Sec.~\ref{sec:washout}, the washout rate is suppressed by the minimum of several rates (see Ref.~\cite{Co:2021qgl} for details), so it is not always straightforward to have washout occur early enough. For gravity mediation with $m_S \sim m_\lambda$, we find that because of the suppression by the $R$ symmetry breaking by the gluino mass, $S_{\rm wo} > 10 f_a$ is possible only for $f_a$ close to the astrophysical lower bound of $10^8$ GeV. $S_{\rm wo} > 100 f_a$ requires additional $R$ symmetry breaking beyond the MSSM. These strong requirements may be avoided if $m_S \ll m_\lambda$ as in gauge mediation, but that requires more analysis (e.g., how the PQ symmetry is spontaneously broken in gauge mediation~\cite{Arkani-Hamed:1998mzz,Asaka:1998ns,Asaka:1998xa}), and is beyond the scope of the present paper.

The parametric resonance discussed in Sec.~\ref{sec:Qball_PR} may also restore the PQ symmetry and solve the domain-wall and isocurvature problems. Even if the PQ symmetry is restored after parametric resonance in the sense that the spatial average of $P$ is zero, $P$ is not fixed to the origin (unlike the thermal trapping case), and $\delta \theta$ remains, and one might worry that the domain wall problem persists.  However, cosmic strings can form~\cite{Kofman:1995fi,Tkachev:1995md,Kasuya:1996ns,Kasuya:1997ha,Kasuya:1998td,Tkachev:1998dc,Kasuya:1999hy}. As we demonstrate in Appendix~\ref{sec:string}, solutions for cosmic strings with non-zero $n_\theta$ exist if the potential is steeper than the quadratic one. This is indeed the case for the vacuum potential of the PQ field. Thus, we expect that cosmic strings form as long as the  PQ field potential is dominated by the vacuum one when parametric resonance occurs and the PQ symmetry is restored.
Consequently, domain walls will have boundaries made of cosmic strings and so can decay as long as the domain wall number is unity. Another way to see the absence of the domain wall problem in the presence of the strings is that at the singularities at the string cores $(P=0)$, $\delta \theta$ is no longer well-defined. The long-wavelength isocurvature perturbations of the misalignment axion dark matter are also absent, because the misalignment abundance is determined by the randomized $\theta$ that results following the symmetry restoration, and is no longer sensitive to the $\delta \theta$ created by the rotation.
For the one-field model in Eq.~\eqref{eq:dim_trans}, we expect that the symmetry is indeed restored by parametric resonance, and therefore the model is free of the domain wall and isocurvature problems.  For the two-field model in Eq.~\eqref{eq:two_field}, the PQ symmetry is not restored as the two fields are strongly fixed at $P\bar{P}= v_P^2$ where the PQ symmetry is broken. The domain walls do not obtain boundaries, and the domain wall problem persists.

Finally, it may be possible to restore the PQ symmetry by Q-ball formation. As we discussed in Sec.~\ref{sec:Qball_PR}, Q-balls can form when the potential of the PQ field is flatter than the quadratic one. This is indeed the case when the potential is dominated by the thermal contribution. For such a potential, fluctuations around the rotation have tachyonic instability modes and grow, and Q-balls eventually form. It may be that these large fluctuations lead to the restoration of the PQ symmetry as in parametric resonance; whether or not this actually occurs needs to be checked by a lattice computation. Also, after Q-ball formation, almost all of the charge is stored in the Q-balls, and field values of $S$ are small outside the Q-balls. Then thermal trapping can occur and the PQ symmetry can be restored. The PQ symmetry restoration, either by large fluctuations or thermal trapping, can solve the domain wall and isocurvature problems.

\section{Summary and discussion}
\label{sec:Disc}

In this paper, we discussed the generation of the curvature perturbations of the universe from the rotation of a complex field charged under an approximate $U(1)$ symmetry. The scenario predicts an observable amount of local non-Gaussianity.

The rotation can be initiated by the Affleck-Dine mechanism. The angular direction of the complex field remains light during inflation and obtains fluctuations, which leads to the fluctuations of the $U(1)$ charge of the rotation. The rotation is thermalized by its interaction with the thermal bath, and the radial mode is dissipated. On the other hand, the angular mode is thermodynamically stable against dissipation into the thermal bath and can be long-lived, dominating the universe. Explicit $U(1)$ symmetry breaking can eventually wash out the $U(1)$ charge, and the angular mode can be dissipated to create entropy. The fluctuation of the charge then creates the curvature perturbations.

The rotation can also generate the baryon asymmetry of the universe. The baryon asymmetry, however, cannot directly come from the $U(1)$ charge of the rotation as in Affleck-Dine baryogenesis, since the rotation is washed out to generate the entropy of the universe. For the axion, the washout can be partial, but the axion rotation does not carry baryon number. Indirect production is possible both for generic flat directions and the axion; the $U(1)$ charge in the rotation may be transferred into asymmetries in the thermal bath prior to washout, and these asymmetries can be processed into $B-L$ asymmetry via $B-L$ breaking interactions, and the $B-L$ asymmetry is not washed out later. This possibility is discussed in Refs.~\cite{Chiba:2003vp,Takahashi:2003db}, where it is  argued that an interaction that simultaneously breaks $U(1)$ and $B-L$ is required. However, as is discussed in~\cite{Domcke:2020kcp,Co:2020jtv} in the context of axion rotations, such a special interaction is actually unnecessary, and $B-L$ production occurs for generic $B-L$ breaking interactions, such as dimension-5 Majorana neutrino mass terms~\cite{Domcke:2020kcp,Co:2020jtv} and $R$-parity violation~\cite{Co:2021qgl}. As long as $B-L$ production dominantly occurs after the entropy production from the washout of the rotation begins, the fluctuation of $n_B/s$ is absent, so no isocurvature perturbations are produced. This is in contrast with the case where the angular mode is assumed to have the same lifetime as the radial mode (which is unlikely as we argued), for which $n_B/s$ will fluctuate.

For the axion rotation, it is also possible to produce the baryon asymmetry by the weak anomaly~\cite{Co:2019wyp}, since the washout of the rotation can be only partial, and the rotation can continue even around the electroweak phase transition. Although only $B+L$ is produced by the weak sphaleron transition~\cite{Klinkhamer:1984di,Kuzmin:1985mm} and baryon asymmetry produced at high temperatures is washed out, a baryon asymmetry proportional to the axion velocity around the electroweak phase transition remains. For the QCD axion with the standard electroweak phase transition temperature $\sim 100$ GeV, the required axion velocity is so large that the kinetic misalignment mechanism overproduces axion dark matter. Successful baryongenesis is possible for a larger electroweak phase transition temperature~\cite{Co:2019wyp} or axion-like particles~\cite{Co:2020xlh}. Such a baryon asymmetry would be produced much after the entropy production, and the isocurvature perturbation is absent.

The rotation can also produce dark matter via the kinetic misalignment mechanism~\cite{Co:2019jts}, where the kinetic energy of the axion rotation is transferred into the axion dark matter density. The production occurs around temperatures of a GeV, long after entropy is produced, so no isocurvature perturbations are produced at the long-wavelength scales that are relevant for the CMB. The axion case without complete washout, however, requires parametric resonance at the early stage of the rotation as is discussed in Sec.~\ref{sec:fluctuations}. The parametric resonance also produces axion dark matter whose abundance is comparable to or larger than that from the kinetic misalignment mechanism~\cite{Co:2017mop,Co:2020dya,Co:2020jtv}. Since this occurs before the entropy production, dark matter isocurvature is produced. In order for the kinetic misalignment mechanism to produce dominant component of dark matter without too large isocurvature perturbations, axions from the parametric resonance must be dissipated, e.g., by thermalization. 

Dark matter and baryon asymmetry do not have to come from the rotation. As long as they are produced after the entropy is produced by the (partial) washout of the rotation, matter isocurvature perturbations are not produced. Viable examples include the freeze-out production of the LSP dark matter around the TeV scale and electroweak baryogenesis.

As we have demonstrated in this paper, rotations of complex fields can produce the curvature perturbations of the universe.  The  thermodynamic stability, which allows the rotation to persist, was a crucial ingredient for the success of the scenario. The persistence of the rotation also opens the possibility that the rotation can have impacts throughout the history of the Universe. As discussed above, because the axion rotation can survive to temperatures of $100\mathchar`-1000$ GeV, it can produce the baryon asymmetry via axiogenesis with weak sphaleron processes.  It can even survive to temperatures of a GeV, in which case it is relevant for the production of axion dark matter via kinetic misalignment. It will be interesting to investigate further possible cosmological roles of rotating fields.

\section*{Acknowledgements}
The work was supported by the DoE office of science under grant DE-SC0011842 at the University of Minnesota (R.C.) and DE-SC0007859 (A.P.) at the University of Michigan.

\appendix

\section{Thermalization of MSSM flat directions}
\label{sec:thermalization}

In this appendix, we discuss the thermalization of the rotation for some examples of MSSM flat directions.

First, let us consider a $\bar{u}_i\bar{u}_j\bar{d}_k \bar{e}_\ell$ flat direction that receives a large field value during inflation. The subscripts are  generation indices with $i\neq j$. The large $S$ field value breaks $SU(3)_c\times U(1)_Y$, but $SU(2)_L$ remains unbroken, so $SU(2)_L$ charged particles and gauge bosons interact efficiently and form a thermal bath after inflation. This flat direction has Yukawa couplings to $Q$, $L$, $H_u$, and $H_d$. If a Yukawa coupling $y$ is large and $y S >T$, the particles that couple to $S$ via the Yukawa coupling
have exponentially small thermal abundance,
but $S$ has one-loop suppressed coupling to the $SU(2)_L$ gauge boson. If a Yukawa coupling is small and $y S <T$, $S$ may be thermalized via the Yukawa interaction. 

\subsection{Thermalization via $SU(2)_L$ gauge bosons}

We first discuss the thermalization by the $SU(2)_L$ gauge boson, whose rate is~\cite{Bodeker:2006ij,Mukaida:2012qn}
\begin{align}
\label{eq:Gammath}
\Gamma_{{\rm th},W} \simeq 10^{-5} \frac{T^3}{S^2}.
\end{align}
Assuming a large enough initial value of $S_i$, the thermalization occurs after the flat direction dominates the energy density of the universe. Then just after the thermalization, $m_S^2 S^2 \simeq \pi^2 g_* T_{\rm th}^4 /30$ and $\Gamma_{{\rm th},W} \simeq 3H \simeq 3 (\pi^2 g_* T_{\rm th}^4 /90)^{1/2}/\MPl$, so
\begin{align}
\label{eq:Tth_Sth}
    T_{\rm th} & \simeq 7 \times 10^{6} \GeV \left(\frac{m_S}{100\TeV}\right)^{ \scalebox{1.01}{$\frac{2}{3}$} } 
    \left(\frac{150}{g_*(T_{\rm th})}\right)^{ \scalebox{1.01}{$\frac{1}{2}$} } ,
    \\ 
    S_{\rm th} & \simeq 4 \times 10^{9} \GeV \left(\frac{m_S}{100\TeV}\right)^{ \scalebox{1.01}{$\frac{1}{3}$} }
    \left(\frac{150}{g_*(T_{\rm th})}\right)^{ \scalebox{1.01}{$\frac{1}{2}$} }.
\end{align}
Note that $S_{\rm th} \gg T_{\rm th}$, so the $U(1)$ charge can indeed remain mostly in the circular rotation after thermalization. Henceforth, we use the number of degrees of freedom $g_*=150$ as a reference point, as the flat direction gives large masses to $SU(3)_c\times U(1)_Y$ gauge multiplets.

If the ellipticity of the initial rotation is $\mathcal{O}(1)$, which is the case in gravity mediation, the circular rotation remaining after thermalization has a comparable energy density to radiation, so the domination by the circular rotation follows immediately after the completion of thermalization. If the washout rate of the rotation is smaller than the thermalization rate in Eq.~(\ref{eq:Gammath}),  washout occurs long after the domination by the circular rotation, and our computation of the curvature perturbation is applicable. This is expected to hold because effective washout requires multiple symmetry breakings, as discussed in Sec.~\ref{sec:washout}.

There are several consistency conditions for our analysis on the thermalization by the $SU(2)_L$ gauge boson to be applicable. First, we  assumed $ yS_{\rm th} > T_{\rm th}$.  This is satisfied when
\begin{align}
\label{eq:mS_y}
    m_S \lesssim 10^7 \GeV \left(\frac{y}{0.01}\right)^3.
\end{align}
Since at least one of $\bar{u}_i\bar{u}_j$ must be a charm or top, this condition is always satisfied for $m_S < 10^7$ GeV. 
Second, we assumed that the rotation is initiated by the zero-temperature potential.
However, the coupling of the flat direction with the weak gauge boson generates a two-loop suppressed thermal-log potential~\cite{Anisimov:2000wx}
\begin{align}
\label{eq:thermal_log}
V_{\rm th} \simeq \alpha_2^2 T^4 \ln\left(\frac{y^2 S^2}{T^2} \right),
\end{align}
and we will use $\alpha_2 = 1/20$ in what follows. To ensure that rotation is initiated by the zero-temperature potential rather than the thermal one requires
\begin{align}
\label{eq:osc}
    S_i \gtrsim 
    \begin{cases}
    10^{16}\GeV 
    \left(\frac{150}{g_*}\right)^{ \scalebox{0.99}{$\frac{1}{2}$} }
    & 
    : T_R \gtrsim 10^{11} \GeV \left(\frac{m_S}{100 \TeV}\right)^{ \scalebox{0.9}{$\frac{1}{2}$} }
    \left(\frac{150}{g_*}\right)^{ \scalebox{0.9}{$\frac{1}{4}$} }
    \\
    10^{15}\GeV \left(\frac{T_R}{10^{10}\GeV}\right) \left(\frac{100 \TeV}{m_S}\right)^{ \scalebox{0.99}{$\frac{1}{2}$} }  
    \left(\frac{150}{g_*}\right)^{ \scalebox{0.99}{$\frac{1}{4}$} } 
    & 
    : T_R \lesssim 10^{11} \GeV \left(\frac{m_S}{100 \TeV}\right)^{ \scalebox{0.9}{$\frac{1}{2}$} }
    \left(\frac{150}{g_*}\right)^{ \scalebox{0.9}{$\frac{1}{4}$} }
    \end{cases},
\end{align}
where $S_i$ is the field value of $S$ when $3H \simeq m_S$ and $T_R$ is the reheat temperature after inflation. The first line corresponds to the initiation of the rotation during radiation domination, while the second line is during matter domination by the inflaton. This condition restricts $M$ and/or $n$ to be sufficiently large; see Eq.~(\ref{eq:Pafter}). For example, for $m_S=10^{5}$ GeV, $n=7$ and $M= \MPl$ fulfills this condition. If this condition is violated, the kick in the angular direction becomes relatively weak. Then, the circular rotation  comes to dominate only long after the completion of the thermalization, and there is a danger that washout will happen first.  In this case, the rotation cannot be the curvaton. Finally, we have assumed the initial rotation dominates the universe before thermalization.  This occurs
if 
\begin{align}
\label{eq:dom}
    S_i \gtrsim 
    \begin{cases}
       10^{16}\GeV \left(\frac{m_S}{100 \TeV}\right)^{ \scalebox{0.9}{$\frac{1}{12}$} }  
       \left(\frac{150}{g_*}\right)^{ \scalebox{0.9}{$\frac{1}{8}$} }
       & 
    : T_R \gtrsim 10^{11} \GeV \left(\frac{m_S}{100 \TeV}\right)^{ \scalebox{0.9}{$\frac{1}{2}$} }
    \left(\frac{150}{g_*}\right)^{ \scalebox{0.9}{$\frac{1}{4}$} } 
    \\
     4 \times 10^{16} \GeV 
     \left(\frac{m_S}{100 \TeV}\right)^{ \scalebox{0.9}{$\frac{1}{3}$} } \left(\frac{10^{10}\GeV}{T_R}\right)^{ \scalebox{0.9}{$\frac{1}{2}$} }  
     \left(\frac{150}{g_*}\right)^{ \scalebox{0.9}{$\frac{1}{4}$} }
     & 
    : T_R \lesssim 10^{11} \GeV \left(\frac{m_S}{100 \TeV}\right)^{ \scalebox{0.9}{$\frac{1}{2}$} }
    \left(\frac{150}{g_*}\right)^{ \scalebox{0.9}{$\frac{1}{4}$} }
    \end{cases}.
\end{align}

For a low reheat temperature, it is possible that Eq.~(\ref{eq:osc}) is satisfied while Eq.~(\ref{eq:dom}) is not.  In this case, while the rotation is initiated by the vacuum potential and so the kick to the angular direction is unsuppressed,  thermalization occurs during radiation domination.  This means the remaining circular rotation can dominate the universe at a temperature $T_{\rm RM}$ only long after the completion of the thermalization, and again there is the danger that washout will occur prior to $T_{\rm RM}$. The charge yield of the rotation is given by the second case of Eq.~(\ref{eq:Ytheta}) with $a_{\cancel{U(1)}} = \mathcal{O}(m_{3/2}/m_S)$ and $T_{\rm RM}$ is given by Eq.~(\ref{eq:TRM}). For $Y_\theta<0.1$, however, Eq.~(\ref{eq:TRM}) gives $S_{\rm RM} < T_{\rm RM}$.  This means matter domination by the rotation actually does not occur, since the rotation is no longer stable for $S<T$; almost all of the charge in the rotation is transferred into the particle-antiparticle asymmetry in the bath, and the rotation dissipates before it has a chance to dominate the energy density.%
\footnote{This is true for MSSM flat directions. However, for generic scalar fields, it is a logical possibility that after the rotation thermalizes, it decouples from the thermal bath again. In this case, the rotation can be stable even if $S<T$ so can come to dominate the universe. Once the rotation recouples to the bath afterward, the entropy of the universe can be dominantly created from the rotation.}
In this case, the rotation cannot act as a successful curvaton, since the non-Gaussianity of the curvature perturbations becomes too large.

\subsection{Thermalization via Yukawa couplings}

We next consider the thermalization by a Yukawa coupling, whose rate is~\cite{Mukaida:2012qn}
\begin{align}
    \Gamma_{{\rm th},y} \simeq \alpha_2 y^2 T,
\end{align}
and we use $\alpha_2 = 1/30$ in the following. Assuming the thermalization after the flat direction dominates the energy density of the universe,
\begin{align}
    T_{\rm th} & \simeq 6 \times 10^6 \GeV \left(\frac{y}{3 \times 10^{-5}}\right)^2
    \left(\frac{150}{g_*(T_{\rm th})}\right)^{ \scalebox{1.01}{$\frac{1}{2}$} },
    \\
    S_{\rm th} & \simeq 2 \times 10^9 \GeV 
    \left(\frac{y}{3 \times 10^{-5}}\right)^4
    \left(\frac{100 \TeV}{m_S}\right)
    \left(\frac{150}{g_*(T_{\rm th})}\right)^{ \scalebox{1.01}{$\frac{1}{2}$} }.
\end{align}
$S_{\rm th} \gg T_{\rm th}$ is easily satisfied.

The condition $y S_{\rm th} < T_{\rm th}$ requires that
\begin{align}
\label{eq:yup}
    y \lesssim 10^{-4} \left(\frac{m_S}{100 \TeV}\right)^{ \scalebox{0.9}{$\frac{1}{3}$} }.
\end{align}
This condition must be satisfied for max$(y_{u_i},y_{u_j})$ or max$(y_{d_k},y_{e_\ell})$, since they give masses to $H_u$ and $H_d$, respectively. max$(y_{u_i},y_{u_j}) \geq y_c$ violates this condition for realistic $m_S$, and the thermalization by the up-type Yukawa coupling cannot occur.

The thermalization by $y$ is more effective than that by the one-loop suppressed coupling with the $SU(2)_L$ gauge boson if
\begin{align}
\label{eq:ylow}
    y \gtrsim 3 \times 10^{-5} \left(\frac{m_S}{100 \TeV}\right)^{ \scalebox{1.01}{$\frac{1}{3}$} }.
\end{align}
For $m_S = 100$--$1000$ TeV, the observed Higgs mass requires that ${\rm tan}\beta \lsim 3$, so the conditions in Eqs.~(\ref{eq:yup}) and (\ref{eq:ylow}) can be satisfied only by $y_{d}$. For $1 \TeV < m_S \lesssim 100$ TeV, ${\rm tan}\beta$ can be larger and these two conditions can be satisfied by $y_d$ and $y_e$.

The Yukawa coupling can also impact the initiation of the rotation, by giving a one-loop thermal mass $\sim y T$ when the rotation is initiated. The thermal potential is ineffective if one of the following two conditions are met for both max$(y_{u_i},y_{u_j})$ and max$(y_{d_k},y_{e_\ell})$.
\begin{align}
    S_i \gtrsim &
    \begin{cases}
     10^{16} \GeV  
     \left(\frac{10^{-5}}{y}\right)
     \left(\frac{m_S}{100 \TeV}\right)^{ \scalebox{0.9}{$\frac{1}{2}$} } 
     \left(\frac{150}{g_*}\right)^{ \scalebox{0.9}{$\frac{1}{4}$} }
     & 
    : T_R \gtrsim 10^{11} \GeV \left(\frac{m_S}{100 \TeV}\right)^{ \scalebox{0.9}{$\frac{1}{2}$} }
    \left(\frac{150}{g_*}\right)^{ \scalebox{0.9}{$\frac{1}{4}$} } 
    \\
    4 \times 10^{15}\GeV 
    \left(\frac{10^{-5}}{y}\right)
    \left(\frac{m_S}{100 \TeV}\right)^{ \scalebox{0.9}{$\frac{1}{4}$} } \left(\frac{T_R}{10^{10}\GeV}\right)^{ \scalebox{0.9}{$\frac{1}{2}$} } 
    \left(\frac{150}{g_*}\right)^{ \scalebox{0.9}{$\frac{1}{8}$} }
    & 
    : T_R \lesssim 10^{11} \GeV \left(\frac{m_S}{100 \TeV}\right)^{ \scalebox{0.9}{$\frac{1}{2}$} }
    \left(\frac{150}{g_*}\right)^{ \scalebox{0.9}{$\frac{1}{4}$} }
    \end{cases} \nonumber \\
    m_S \gtrsim &
    \begin{cases}
    2\times 10^7\GeV 
    \left(\frac{y}{10^{-5}}\right)^2  
    \left(\frac{150}{g_*}\right)^{ \scalebox{0.9}{$\frac{1}{2}$} }
    &
    : T_R \gtrsim 2 \times 10^{12} \GeV 
    \left( \frac{y}{10^{-5}} \right)
    \left(\frac{150}{g_*}\right)^{ \scalebox{0.9}{$\frac{1}{2}$} }
    \\ 
    600 \TeV 
    \left(\frac{y}{10^{-5}}\right)^{ \scalebox{0.9}{$\frac{4}{3}$} } 
    \left(\frac{T_R}{10^{10}\GeV}\right)^{ \scalebox{0.9}{$\frac{2}{3}$} }
    \left(\frac{150}{g_*}\right)^{ \scalebox{0.9}{$\frac{1}{6}$} }
    &
    : T_R \lesssim 2 \times 10^{12} \GeV 
    \left( \frac{y}{10^{-5}} \right)
    \left(\frac{150}{g_*}\right)^{ \scalebox{0.9}{$\frac{1}{2}$} }
    \end{cases} .
\end{align}
The condition on $S_i$ comes from the case where the particles have masses above $T$, $y S_i > T_{\rm rot}$, so the thermal potential is exponentially suppressed. Here $T_{\rm rot}$ is the temperature at which the rotation is initiated by the vacuum potential, i.e., $m_S = 3H(T_{\rm rot})$. The condition on $m_S$ is when the thermal mass is present but negligible compared to the vacuum mass, $y T_{\rm rot} < m_S$. 

The condition for the flat direction to dominate the energy of the universe before the completion of the thermalization is given by
\begin{align}
    S_i \gtrsim
    \begin{cases}
    9 \times 10^{15} \GeV 
    \left( \frac{y}{3 \times10^{-5}}\right) 
    \left(\frac{100 \TeV}{m_S}\right)^{ \scalebox{0.9}{$\frac{1}{4}$} } 
    \left(\frac{150}{g_*}\right)^{ \scalebox{0.9}{$\frac{1}{8}$} }
    & 
    : T_R \gtrsim 10^{11} \GeV \left(\frac{m_S}{100 \TeV}\right)^{ \scalebox{0.9}{$\frac{1}{2}$} }
    \left(\frac{150}{g_*}\right)^{ \scalebox{0.9}{$\frac{1}{4}$} }
    \\
    3 \times 10^{16} \GeV 
    \left( \frac{y}{3 \times10^{-5}}\right) 
    \left(\frac{10^{10} \GeV}{T_R}\right)^{ \scalebox{0.9}{$\frac{1}{2}$} } 
    \left(\frac{150}{g_*}\right)^{ \scalebox{0.9}{$\frac{1}{4}$} }
    & 
    : T_R \lesssim 10^{11} \GeV \left(\frac{m_S}{100 \TeV}\right)^{ \scalebox{0.9}{$\frac{1}{2}$} }
    \left(\frac{150}{g_*}\right)^{ \scalebox{0.9}{$\frac{1}{4}$} }
    \end{cases}
\end{align}
If this condition is violated, the charge yield is given by Eq.~(\ref{eq:Ytheta}) and the circular rotation dominates the universe at the temperature given by Eq.~(\ref{eq:TRM}).

Based on the above computation, one can see that the circular rotation can indeed dominate the universe for sufficiently large $S_i$ and/or $T_R$. A similar analysis is applicable to $QQQL$, $Q\bar{u} L \bar{e}$, and $Q\bar{u} Q\bar{d}$ flat directions, which keep a $U(1)$, $SU(2)$, and $U(1)$ gauge symmetry unbroken, respectively. The $SU(2)_L$ gauge symmetry in the above analysis should be replaced by the unbroken gauge symmetry. All of these directions have $B-L=0$, so can be washed out by the interactions and masses in the MSSM.

\subsection{Q-balls}

In the above discussion, we neglected the possible production of Q-balls. Since the thermal potential is flatter than the quadratic one, when it dominates over the vacuum potential, Q-balls can form~\cite{Kasuya:2001hg,Kasuya:2010vq}. However, we argue that those Q-balls disappear before the circular rotation dominates the universe. The condition for a Q-ball solution to exist is that $V(S,T)/S^2$ has a global minimum at $S_0\neq 0$~\cite{Coleman:1985ki}. It is also required that the distance between Q-balls should be larger then the radii of Q-balls~\cite{Chiba:2010ff}; otherwise Q-balls coalesce and then the field configuration reverts to a nearly homogeneous rotation. This condition is equivalent to requiring that the charge density inside Q-balls $\sim m_S S_0^2$ be larger than the average charge density of the universe $n_\theta$. In our setup, the latter is violated first; in order to serve as a curvaton, the rotation must dominate the universe, so $n_\theta$ is quite large.

First we consider the case with $y S >T$.
Then the combination of the thermal log potential in Eq.~(\ref{eq:thermal_log}) and the zero temperature potential (with correction $K m_S^2|P|^2 {\rm ln}|P|^2$) gives a global minimum for $V(S,T)/S^2$ at an $S_0$ given by $K m_S^2 S_0^2 \sim \alpha_2^2 T^4$. The Q-balls disappear when
\begin{align}
  \frac{m_S S_0^2}{n_\theta} \sim  \frac{\alpha_2^2 T^4}{K m_S n_\theta} \sim \frac{10^{-5}}{K} \frac{T}{T_{\rm RM}} <1.
\end{align}
In the second equality, we have replaced $n_\theta$ in favor of the temperature $T_{\rm RM}$ where the rotation dominates using Eq.~(\ref{eq:TRM}).
As we discussed in Sec.~\ref{sec:Qball_PR}, the two-loop corrections from the gauge interactions and scalar masses generate $K\sim 10^{-3}$, so the Q-balls disappear well before the circular rotation dominates at $T_{\rm RM}$.

We next consider the case with $y S <T$. The one-loop thermal potential induced by the Yukawa coupling is, in the high temperature expansion,
\begin{align}
  V(S,T) \simeq \frac{1}{24}y^2 S^2 T^2 - \frac{1}{12\pi}y^3 S^3 T + \cdots.
\end{align}
Then, the location of the minimum $S_0$ is found where $y^3 T S_0/(12\pi) \sim K m_S^2$.
Then Q-balls disappear when
\begin{align}
    \frac{m_S S_0^2}{n_\theta} \sim \frac{(12\pi K)^2 m_S^5}{y^6 T^2 n_\theta} < \frac{T^4}{12\pi K m_S n_\theta} \sim \frac{10^{-3}}{K} \frac{T}{T_{\rm RM}} <1
\end{align}
where we imposed $ y S_0 < T $ in the first inequality, so that the one-loop thermal potential is not exponentially suppressed. Even for this optimal Yukawa coupling, the Q-balls disappear before the circular rotation dominates.

\section{Numerical computation of the charge fluctuation}
\label{sec:charge}
In this Appendix, we compute the dependence of the charge density on the initial angle $\theta_i$. The equation of motion of the field $P$ is
\begin{align}
    \frac{{\rm d}^2P}{{\rm d}t^2} + 3 H \frac{{\rm d}P}{{\rm d}t} + \frac{\partial V}{\partial P^*} = 0.
\end{align}
We make the time and field value dimensionless by the following change of variables,
\begin{align}
    t = \frac{1}{m_S}z,~~
    P = \left[\frac{m_S M^{n-2}}{n^{1/2} (n+1)}\right]^\frac{1}{n-1} p.
\end{align}
Using the potential in Eq.~\eqref{eq:potential}, the resulting equation of motion is given by
\begin{align}
\label{eq:newvars}
    \ddot{p} + \frac{2}{(1+w)z} \dot{p} + p |p|^{2n-2} + a_{\cancel{U(1)}} \frac{n-2}{n^{1/2}} p^{*n} + p - c_H\frac{4}{9(1+w)^2 z^2} p = 0,
\end{align}
where the dots represent derivatives with respect to $z$. Here, $\omega = 1/3$ and $0$ for radiation domination and matter domination, respectively. When $z \ll 1$, the fourth and fifth terms are negligible and $p$ follows an attractor solution if $n>3$ for matter domination and $n>5$ for radiation domination~\cite{Dine:1995kz,Harigaya:2015hha}:
\begin{align}
\label{eq:attractor}
    |p(z)| = \left(\left(\frac{4 c_H}{9(1+w)^2} + \frac{2}{(n-1)(1+w)} - \frac{n}{(n-1)^2}\right) \frac{1}{z^2}\right)^\frac{1}{2n-2}.
\end{align}
One can show that Eq.~(\ref{eq:attractor}) is indeed an attractor by the change of variables $p =  z^{-1/(n-1)} \times \psi$ and $z \propto {\rm exp}(3(1+w)N/2)$. Then the potential terms for $\psi$ as well as its minimum are $N$-independent, and the sign of the friction term of the equation of motion of $\psi$ is positive if $n > 1 + 2(1+w)/(1-w)$.

%%%
\begin{figure}
\includegraphics[width=0.9\linewidth]{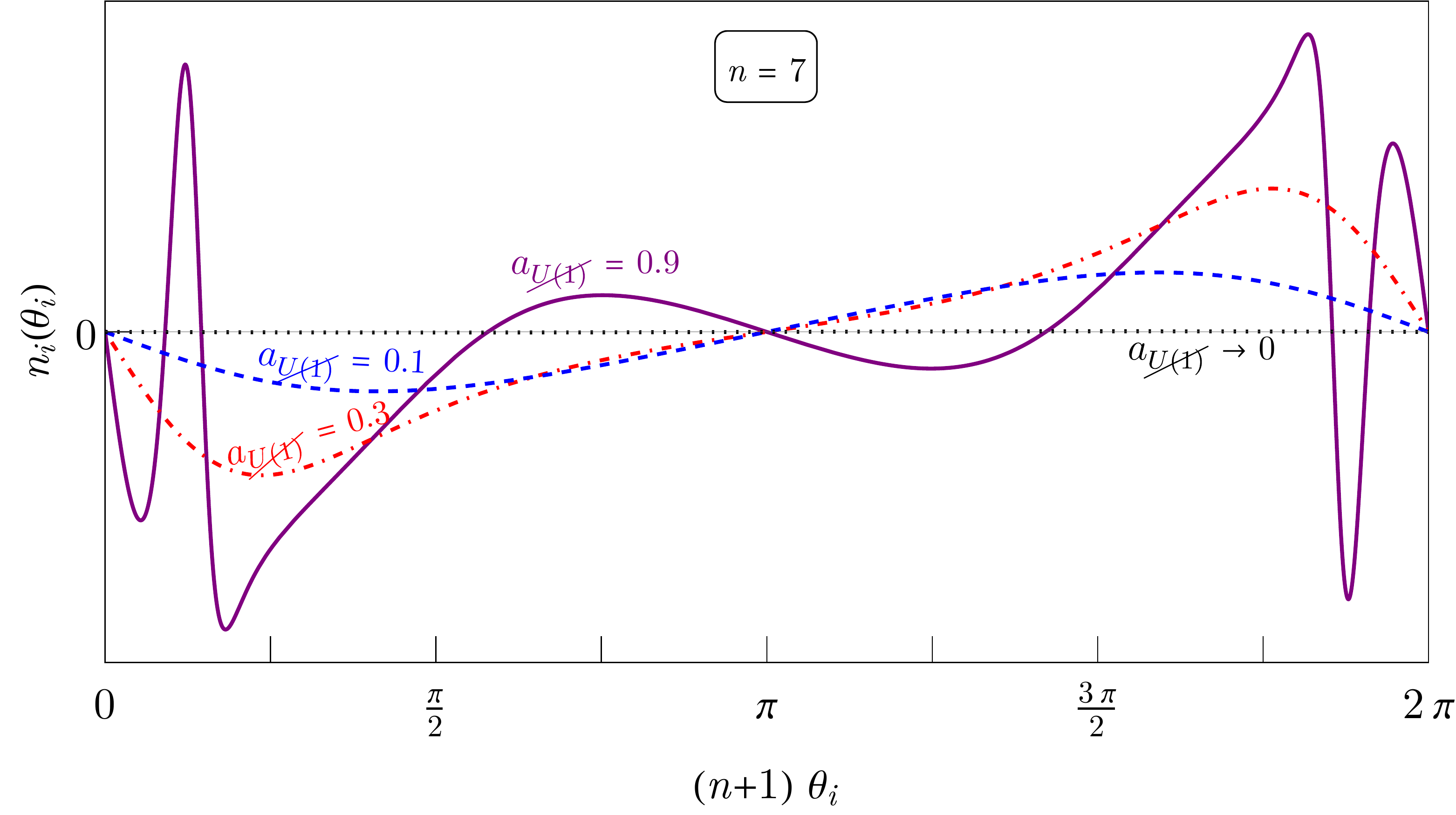}
\caption{The dependence of the charge density on the initial angle $\theta_i$ for $n=7$ and several values of the $U(1)$ breaking parameter $a_{\cancel{U(1)}}$ defined in Eq.~(\ref{eq:potential}).}
\label{fig:charge}
\end{figure}
%%%

We numerically solve Eq.~\eqref{eq:newvars} starting from $z= z_0 \ll 1$ with an initial condition following the attractor solution,
\begin{align}
    p(z_0) = &  \left(\left(\frac{4 c_H}{9(1+w)^2} + \frac{2}{(n-1)(1+w)} - \frac{n}{(n-1)^2}\right) \frac{1}{z_0^2}\right)^\frac{1}{2n-2} \times {\rm exp}(i \theta_i),~~0 \leq \theta_i \leq \frac{2\pi}{n+1}, \nonumber \\
    p'(z_0) =& - \frac{1}{(n-1)z_0} \times p(z_0),
\end{align}
and compute the charge density $\propto i (p^*p'- p p^{*'})$ as a function of $\theta_i$ at a fixed time $z_i \gg 1$ such that the explicit PQ breaking is negligible. This is the case for $z_i \gtrsim 10$. As long as the universe is dominated by the inflaton or the radiation created from it at $z= z_i$, we may identify $z_i$ as a flat time slice.

In Fig.~\ref{fig:charge}, we show the charge density in an arbitrary unit for $n=7$ for several values of $a_{\cancel{U(1)}}$. Here we assume radiation domination. For small $a_{\cancel{U(1)}}$, the sign of the charge is simply determined by the sign of the potential gradient in the angular direction at $\theta_i$ and is negative for $0 < (n+1)\theta_i < \pi$ and positive for $\pi < (n+1)\theta_i < 2\pi$. For $a_{\cancel{U(1)}} = \mathcal{O}(1)$, the angular direction oscillates before the explicit breaking becomes ineffective, and the charge non-trivially depends on $\theta_i$.

\section{Rotating string}
\label{sec:string}
In this Appendix, we demonstrate the existence of cosmic strings that rotate in field space. The rotating string is analogous to vortices that appear in Bose-Einstein condensation in condensed matter systems. The existence of these strings is important for the successful collapse of the domain walls following PQ restoration via parametric resonance, as discussed in Sec.~\ref{sec:fluctuations}.  We consider a rotating configuration with cylindrical symmetry:
\begin{align}
P = \frac{1}{\sqrt{2}}S(r)e^{i\omega t + i m \phi},
\end{align}
with $m$ an integer, $(r,\phi)$ the cylindrical coordinate variables, $S(0) =0$, and $S(\infty)=S_\infty$. The equation of motion gives
\begin{align}
\label{eq:eom_string}
    \frac{{\rm d}^2S}{{\rm d}r^2}  + \frac{1}{r} \frac{{\rm d}S}{{\rm d}r}  - \frac{m^2}{r^2}S + \omega^2 S - V'(S) = 0.
\end{align}
At $r= \infty$, the last two terms dominate, so we obtain a relation between $\omega$ and $S_\infty$,
\begin{align}
\label{eq:omega}
    \omega^2 = \frac{V'(S_\infty)}{S_\infty}.
\end{align}
In the cosmology we consider, we may estimate $\omega$ and $S_\infty$ from this relation for fixed charge density $n_\theta = \omega S_\infty^2$.

We may demonstrate the existence of cosmic string solutions by employing the technique used to show a bounce solution in tunneling problems~\cite{Coleman:1977py} and a Q-ball solution~\cite{Coleman:1985ki}, wherein we regard $S$ as a trajectory with $r$ as a time variable. The equation of motion of $S$ is that of a particle moving in 1D with a  $1/r$ friction term and a ``time"-dependent potential
\begin{align}
    V_{\rm eff}(S) = \frac{1}{2}\left(\omega^2 - \frac{m^2}{r^2}\right)S^2 - V(S).
\end{align}
Around $r \sim 0$, the first three terms in Eq.~(\ref{eq:eom_string}) dominate and $S$ increases in proportion to $ r^{|m|}$,
\begin{align}
    S(r) \rightarrow  b r^{|m|} ~~(r \rightarrow 0),
\end{align}
where $b$ is a constant determined by satisfying the condition $S(\infty)=S_\infty$. At large $r$, the potential $V_{\rm eff}$ becomes $r$-independent. If
\begin{align}
    V_{\rm eff}''(S_\infty) = \omega^2 - V''(S_\infty) = \frac{V'(S_\infty)}{S_\infty}-V''(S_\infty)  <0,
\end{align}
namely, $V(S)$ is steeper than a quadratic potential at $S_\infty$, then $V_{\rm eff}$ has a maximum at $S_\infty$ for $r \rightarrow \infty$. If $b$ is too small, $S$ does not reach $S_\infty$ and goes back to $0$. If $b$ is too large, $S$ goes over the maximum and runs away to infinity. Therefore, there is a value of $b$ such that $S$ lands onto $S_\infty$ at $r = \infty$. This solution corresponds to a cosmic string. The core size is determined by the curvature of $V_{\rm eff}$,
\begin{align}
\label{eq:rc}
    r_c^{-1} \equiv \sqrt{- V''_{\rm eff}(S_\infty)} = \sqrt{V''(S_\infty)-\frac{V'(S_\infty)}{S_\infty}}.
\end{align}
%%

%%%
\begin{figure}
\includegraphics[width=0.8\linewidth]{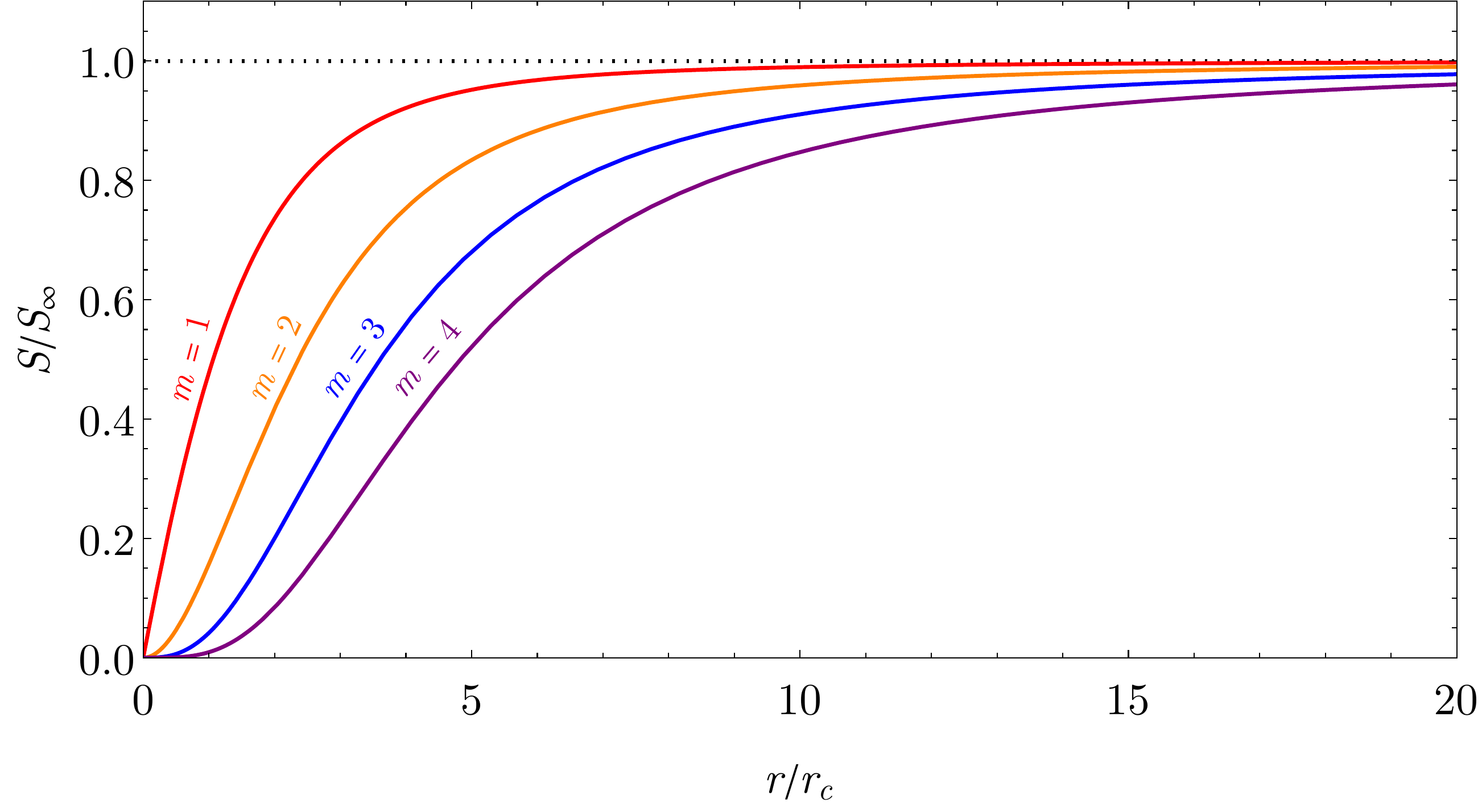}
\caption{Radial profiles of rotating string solutions as a function of azimuthal quantum number $m$ for the potential in Eq.~(\ref{eq:potential_string}).  The radius is normalized by the string core size $r_c$ defined in Eq.~(\ref{eq:rc}), and the field value $S$ is normalized by its asymptotic value $S_0$. Here we take $\Delta = S_{0}/m_{S}=50$, but as discussed in the text, the profiles are nearly independent of this choice for $\Delta \gg 1$.}
\label{fig:string}
\end{figure}
%%%

The condition for the string solution to exist, namely, a potential $V(S)$ steeper than a quadratic one, is complementary to the condition for the Q-ball solution to exist, a potential flatter than a quadratic one~\cite{Coleman:1985ki}. Note also unlike the usual case without rotations, the cosmic string solutions do not require $V(S)$ itself to have a minimum at non-zero $S$; this is because of the centrifugal potential. This allows rotating strings even for the MSSM flat directions. Like non-rotating global strings, the tension of the rotating string is IR-divergent, which is cut off by the distance between strings.

As an example, let us consider the following potential, typical of MSSM flat directions,
\begin{align}
\label{eq:potential_string}
    V(S) = \frac{1}{2}m_S^2 S^2\left( 1 + K \ln \left( \frac{S^2 + m_S^2}{m_S^2} \right) \right).
\end{align}
The logarithmic term arises as a result of quantum correction that has an IR cutoff at the soft mass scale $m_S$.
If $K >0$, the potential is steeper than a quadratic one, and we expect that a rotating cosmic string solution exists.
It is convenient to normalize $S$ by $S_\infty$ and $r$ by $r_c$, i.e., $\sigma = S/S_\infty$  and $\rho = r/r_c$. In these variables, the equation of motion is
\begin{align}
\label{eq:eom_sigma_rho}
    \frac{{\rm d}^2\sigma}{{\rm d}\rho^2}  + \frac{1}{\rho} \frac{{\rm d}\sigma}{{\rm d}\rho} - \frac{m^2}{\rho^2} \sigma  
    + \frac{\left(\Delta ^2+1\right)^2  \left(\frac{1}{\Delta ^2 \sigma
   ^2+1}-\frac{1}{\Delta ^2+1} - \ln \left(\frac{\Delta ^2 \sigma
   ^2+1}{\Delta ^2+1}\right)\right)}{2 \Delta ^2 \left(\Delta
   ^2+2\right)} \sigma  = 0 ,
\end{align}
where $\Delta = S_\infty / m_S$.
Here $\omega$ is chosen according to Eq.~(\ref{eq:omega}). Note that the $K$-dependence disappears from the equation of motion. In Fig.~\ref{fig:string}, we show normalized profiles of the string solutions, where the various curves correspond to different choices of $m$. Here we take $\Delta = 50$, but the normalized profile is nearly independent of $S_\infty/m_S$ as long as $S_\infty \gg m_S$. This is because in the limit $S_\infty \gg m_S$, the last term in Eq.~(\ref{eq:eom_sigma_rho}) simplifies to $-\sigma \ln \sigma$, which is independent of $\Delta$. In this limit, $r_c \simeq 1 / (\sqrt{2K} m_S)$.

\nocite{apsrev41Control}
\bibliographystyle{apsrev4-1}
\bibliography{curvaton}

\end{document}